\documentclass[aps,prd,preprint,showpacs,amsmath,amssymb,floatfix]{revtex4}
\usepackage{epsfig}
\usepackage{enumerate}
\usepackage{subfigure}
\usepackage{rotating}
\usepackage[bookmarks]{hyperref}
\usepackage{color}

\newlength{\www}

\newcommand{\be}{\begin{equation}}
\newcommand{\ee}{\end{equation}}
\newcommand{\ba}{\begin{eqnarray}}
\newcommand{\ea}{\end{eqnarray}}

\newcommand{\bq}{\begin{equation}}
\newcommand{\eq}{\end{equation}}
\newcommand{\bqa}{\begin{eqnarray}}
\newcommand{\eqa}{\end{eqnarray}}
\newcommand{\ben}{\begin{enumerate}}
\newcommand{\een}{\end{enumerate}}
\newcommand{\bc}{\begin{center}}
\newcommand{\ec}{\end{center}}
\newcommand{\bqb}{\begin{eqnarray*}}
\newcommand{\eqb}{\end{eqnarray*}}

\newcommand{\psl}{p\hskip-0.21cm\slash}
\newcommand{\pgsl}{p_g\hskip-0.35cm\slash~}

\newcommand{\esl}{e\hskip-0.21cm\slash}

\begin{document}

\title{\vspace{1cm}
One-loop electroweak effects on \\ stop-chargino production at LHC
}
\author{
M.~Beccaria$^{a,b}$,
G.~Macorini$^{c}$,
E.~Mirabella$^d$,
L.~Panizzi$^{e, f}$,
F.M.~Renard$^g$
and C.~Verzegnassi$^{e, f}$ \\
\vspace{0.4cm}
}

\affiliation{\small
$^a$ $\mbox{Dipartimento di Fisica, Universit\`a del Salento, Italy}$ \\
\vspace{0.2cm}
$^b$ INFN, Sezione di Lecce, Italy\\
\vspace{0.2cm}
$^c$ ICTP, Trieste, Italy\\
\vspace{0.2cm}
$^d$ Max-Planck-Institut f\"ur Physik (Werner Heisenberg Institut),
Germany \\ 
\vspace{0.2cm}
$^e$ $\mbox{Dipartimento di Fisica Teorica, Universit\`a di Trieste, Italy}$ \\
\vspace{0.2cm}
$^f$ INFN, Sezione di Trieste, Italy\\
\vspace{0.2cm}
$^g$ Laboratoire de Physique Th\'{e}orique et Astroparticules,
Universit\'{e} Montpellier II, France
}

\begin{abstract}

The process of stop-chargino production at LHC has been calculated in the
Minimal Supersymmetric Standard Model at the complete electroweak
one-loop level, assuming a mSUGRA symmetry breaking scheme. Several
properties of the angular and invariant mass distributions of the basic
$b~g\to\tilde{t}_a~ \chi^-_i$ amplitudes have been derived. For a meaningful
collection of different benchmark points the overall electroweak one-loop
effects are at most of the order of a few percent. At the realistically
expected LHC accuracy, the main supersymmetric electroweak features of
the process can be therefore essentially derived in this theoretical scheme
from the simple Born level expressions.


\end{abstract}



\maketitle

\section{Introduction}
\label{sec:intro}

The process of associated stop-chargino production at LHC has been
recently considered as a potential source of information on SUSY
parameters. In particular, it has been shown that the total rate would
exhibit a possibly relevant dependence on $\tan\beta$ \cite{Beccaria:2006wz}
and could also be sensitive to possible deviations from a
Minimal Flavor Violation scheme \cite{Bozzi:2007me}. In both cases, the
calculations have been performed at the lowest electroweak order. SUSY
QCD effects have been computed at NLO \cite{Jin:2002-2003}. The conclusion
was that these NLO strong supersymmetric effects in general enhance the
LO total cross sections significantly, and thus must be carefully taken into
account.\newline
If Supersymmetry were discovered at LHC, and measurements of stop-chargino
production began to be performed, the reasonable question would
arise of whether the NLO electroweak supersymmetric effects might
effectively change the special and relevant SUSY parameter dependence
of the lowest order expressions given in Refs.~\cite{Beccaria:2006wz,Bozzi:2007me},
in which case they should be also carefully taken into account, like the NLO QCD component.
The aim of this paper is precisely that of performing an accurate calculation of
the complete one-loop electroweak supersymmetric contributions to the
stop-chargino production process. As a preliminary approach, we shall
work in the Minimal Supersymmetric Standard Model, accept the validity
of a mSUGRA symmetry breaking scheme and select a number of
meaningful ``benchmark'' points to produce the final numerical
predictions.\newline
Technically speaking, the paper is organized as follows. Sect.\ref{sec:kin} will be
devoted to a description of the shape and of the basic properties of the
parton level amplitudes for $b~g\to\tilde{t}_a~ \chi^-_i$ at Born and at one-loop
level.
A detailed analysis at Born level of the dependence of the total rate
on supersymmetric parameters will be performed.
Illustrations will be given for the angular distributions and for the
invariant mass dependence of the helicity amplitudes near threshold and in
the high energy range, where we have checked the agreement with the
logarithmic terms of the Sudakov expansion. Sect.\ref{sec:oneloopresults} will exhibit the
numerical one-loop effects on the production rates for a selected number
of typical SUSY benchmark points. As a general feature, the effects will
turn out to be numerically small, of a relative few percent at most, which
would hardly be effective at the realistically expected LHC experimental
accuracy.

\section{Kinematics and Amplitudes of the process $b~g\to\tilde{t}_a~ \chi^-_i $}
\label{sec:kin}

The kinematics of the process $b~g\to\tilde{t}_a~ \chi^-_i $ 
is expressed in terms of the Dirac spinors $u(p_b,\lambda_b)$
and $\bar v_c(p_{\chi^-_i},\lambda_{\chi^-_i})$ with the
momenta:
\bq
p_b=(E_b;0,0,p)~~~~~~~~
~p_{\tilde{t}_a}=(E_{\tilde{t}_a};p'\sin\theta,0,p'\cos\theta)
\eq
\bq
p_g=(p;0,0,-p)~~~~~~~~
~p_{\chi^-_i}=(E_{\chi^-_i};-p'\sin\theta,0,-p'\cos\theta)
\eq
and the gluon polarization vector:
\bq
e_g(\lambda_g)=(0;{\lambda_g\over\sqrt{2}},-~{i\over\sqrt{2}},0)
\eq
referring to the helicity labels $\lambda_b=\pm1/2$,
$\lambda_g=\pm1$, $\lambda_{\chi^-_i}=\pm1/2$.\\
The angle $\theta$ refers to $p_{\tilde{t}_a}$ and $p_b$.
We will use $s=(p_b+p_g)^2$, $t=(p_b-p_{\tilde{t}_a})^2$
and $u=(p_b-p_{\chi^-_i})^2$.\\
The top squark states $\tilde{t}_a$ $(a=1,2)$ are mixed states of
$\tilde{t}_{L,R}$ with an angle $\theta_t$ and the chargino
states $\chi^-_i$  $(i=1,2)$ are mixed states of gauginos
and Higgsinos with matrix elements $Z^{\pm}_{ij}$.\\
The process will be described by 8 helicity amplitudes 
$F_{\lambda_b,\lambda_g,\lambda_{\chi^-_i}}$ related to the
8 invariant amplitudes ($k=1, 4$ and $\eta=R, L$):
\bq
A=({\lambda\over2})_{b,\tilde{t}_a}
\sum_{k, \eta} \left[\bar v_c(p_{\chi^-_i})J_{k\eta}u(p_b,\lambda_b)\right]\,
 N_{k\eta}(s,t,u)
\eq
\bq
J_{1\eta}=\pgsl\esl P_{\eta}
~~~~~~
J_{2\eta}=(e\cdot p_{\chi}) P_{\eta}
\eq
\bq
J_{3\eta}=\esl P_{\eta}
~~~~~~
J_{4\eta}=(e\cdot p_{\chi})\pgsl P_{\eta}
\eq
with $P_{\eta}=P_{R,L}=(1\pm\gamma^5)/2$. A colour matrix element
$({\lambda\over2})$ relating the initial $b$
quark and the final $\tilde{t}_a$ squark has been 
systematically factorized out.\\
Averaging over initial spins and colours and summing over final spins and colours
with:
\bq
\sum_{col}<{\lambda^l\over2}><{\lambda^l\over2}>=4
\eq
leads to the elementary cross section:
\bq
{d\sigma\over d\cos\theta}= {\beta'\over768\pi s\beta}
\sum_{spins}|F_{\lambda_b,\lambda_g,\lambda_{\chi^-_i}}|^2
\eq
where $\beta=2p/\sqrt{s}$, $\beta'=2p'/\sqrt{s}$. \\
The 8 scalar functions $ N_{k\eta}(s,t,u)$ are obtained 
in terms of Born and one-loop diagrams.\\
The Born terms result from the s-channel b exchange
and the u-channel $\tilde{t}_a$ exchange:
\bq
N^{Born~s}_{1\eta}=-g_s
{A^{\eta}_i(\tilde{t}_a)\over s-m^2_b}
\eq
\bq
N^{Born~u}_{2\eta}=2g_s
{A^{\eta}_i(\tilde{t}_a)\over u-m^2_{\tilde{t}_a}}
\eq
with the couplings:
\bq
\label{couplings}
A^L_i(\tilde t_L^{~})=-~{e\over s_W}Z^+_{1i}~~~~~
A^L_i(\tilde t_R^{~})=
{e m_t\over \sqrt{2}M_Ws_W\sin\beta}Z^+_{2i}~~~~~
A^R_i(\tilde t_L^{~})=
{e m_b\over \sqrt{2}M_Ws_W\cos\beta}Z^{-*}_{2i}
\eq
Using the Dirac decomposition explicitly given in App.~A of \cite{Beccaria:2006wz} one gets
the Born contribution to the 8 helicity amplitudes. Let us notice their
basic properties which will be essential to understand the final results.
First, because of the small value of $m_b$, the $b$ helicity corresponds
to the chirality $\eta$ (L for $\lambda_b=-1/2$ and R for $\lambda_b=+1/2$).
In the case of the production of the lightest chargino ($i=1$) this means that
the $\lambda_b=-1/2$ amplitudes will generally dominate because the R
chirality couplings (see eq.(\ref{couplings})) are depressed by the $m_b$ factor
and by the non-diagonal chargino mixing element $Z^{-*}_{2i}$.
One can then predict the main features of the angular and of the energy
dependences using again App.~A of \cite{Beccaria:2006wz}. At low energy (near above threshold)
the u-channel contribution is suppressed by the final momentum $p'$.
Only the s-channel contribution survives and the leading amplitudes
should be $F_{--+}$ and $F_{---}$. They respectively produce an angular
distribution $(1-\cos\theta)$ and $(1+\cos\theta)$. Having the same magnitude
(at Born level) the unpolarized cross section should then be flat.\\
At high energy $(\sqrt{s} \gg m)$ one observes a cancellation between the
s-channel and the u-channel Born contributions to $F_{+++}$, $F_{++-}$,
$F_{---}$, $F_{--+}$ (see App.A of \cite{Beccaria:2006wz}), as well as the mass suppression
of the u-channel contribution to $F_{+-+}$, $F_{-+-}$ (because $r_b$
and $r_{\chi}$ tend to $1$). The only surviving amplitudes at high
energy are then $F_{+--}$ and $F_{-++}$:
\bq
F^{Born~u}_{+--} \to g_s\sqrt{2}
A^R_i(\tilde{t}_a)\sin{\theta\over2}~~~
F^{Born~u}_{-++} \to -g_s\sqrt{2}
A^L_i(\tilde{t}_a)\sin{\theta\over2}
\label{high_energy_amplitudes}
\eq
In this high energy limit the quantities of Eq.(\ref{high_energy_amplitudes}) 
can be expressed in terms of 3 basic
amplitudes, one of gaugino type $F_{-++}(\tilde t_L^{~})$ and two of
higgsino type $F_{-++}(\tilde t_R^{~})$, $F_{+--}(\tilde t_L^{~})$. In all
cases the high energy distribution should tend to a $(1-\cos\theta)$
shape. For $\chi_1$ production and for the reasons already given
above, $F_{-++}$ should dominate. For a light stop $\tilde t_1^{~}$, mixture
of $\tilde t_L^{~}$ and $\tilde t_R^{~}$, this amplitude will be:
\bq
\cos\theta_t F_{-++}(\tilde t_L^{~}) +\sin\theta_t F_{-++}(\tilde t_R^{~})
\eq

\subsection{Parameter Dependence at Born Level}
\label{sec:pardep}

\begin{figure}
\centering
\begin{minipage}{\textwidth}
  \subfigure[]{
  \epsfig{file=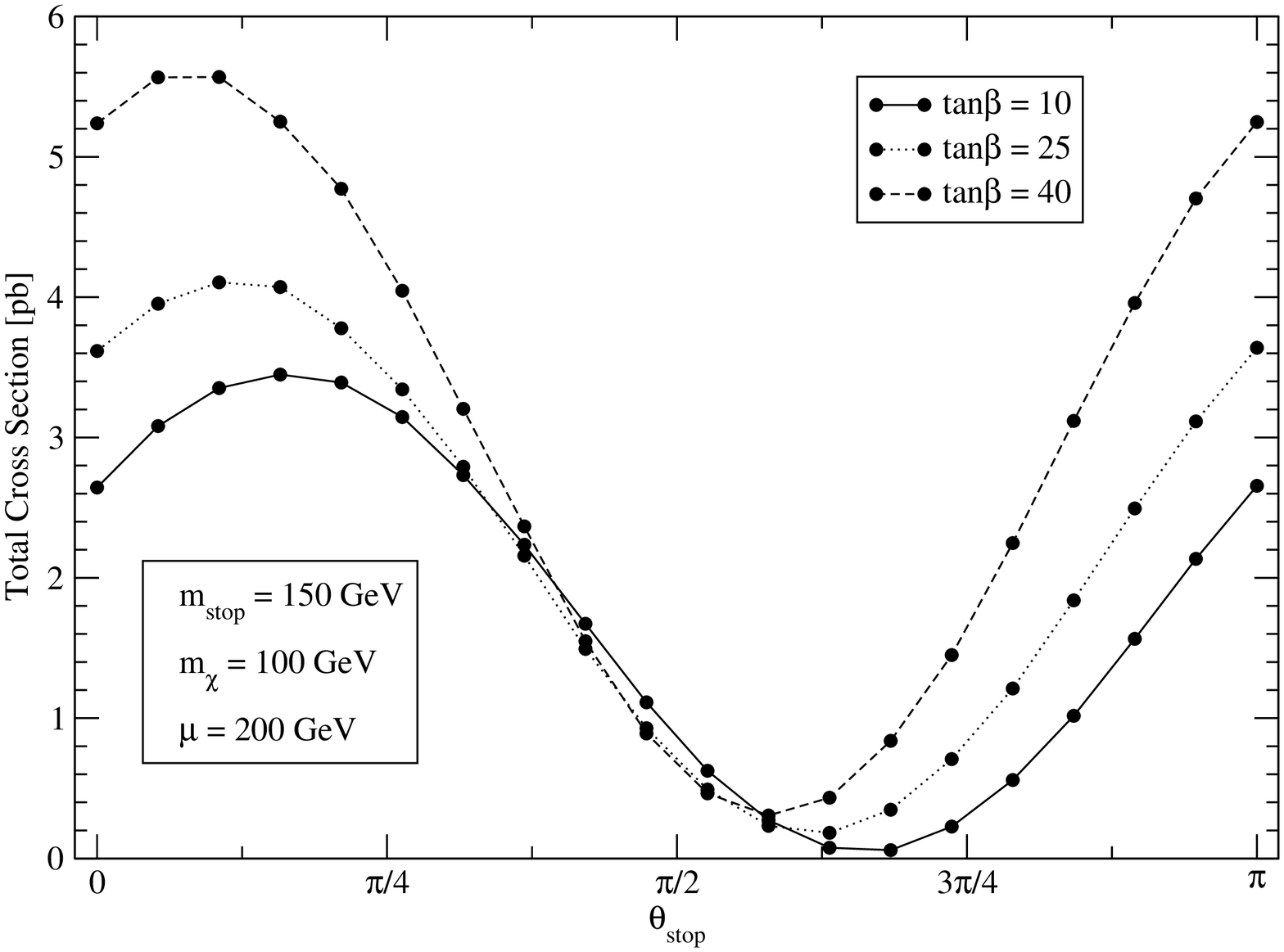, width=0.48\textwidth, angle=0}
  \label{fig:parameter:a}}\hfill
  \subfigure[]{
  \epsfig{file=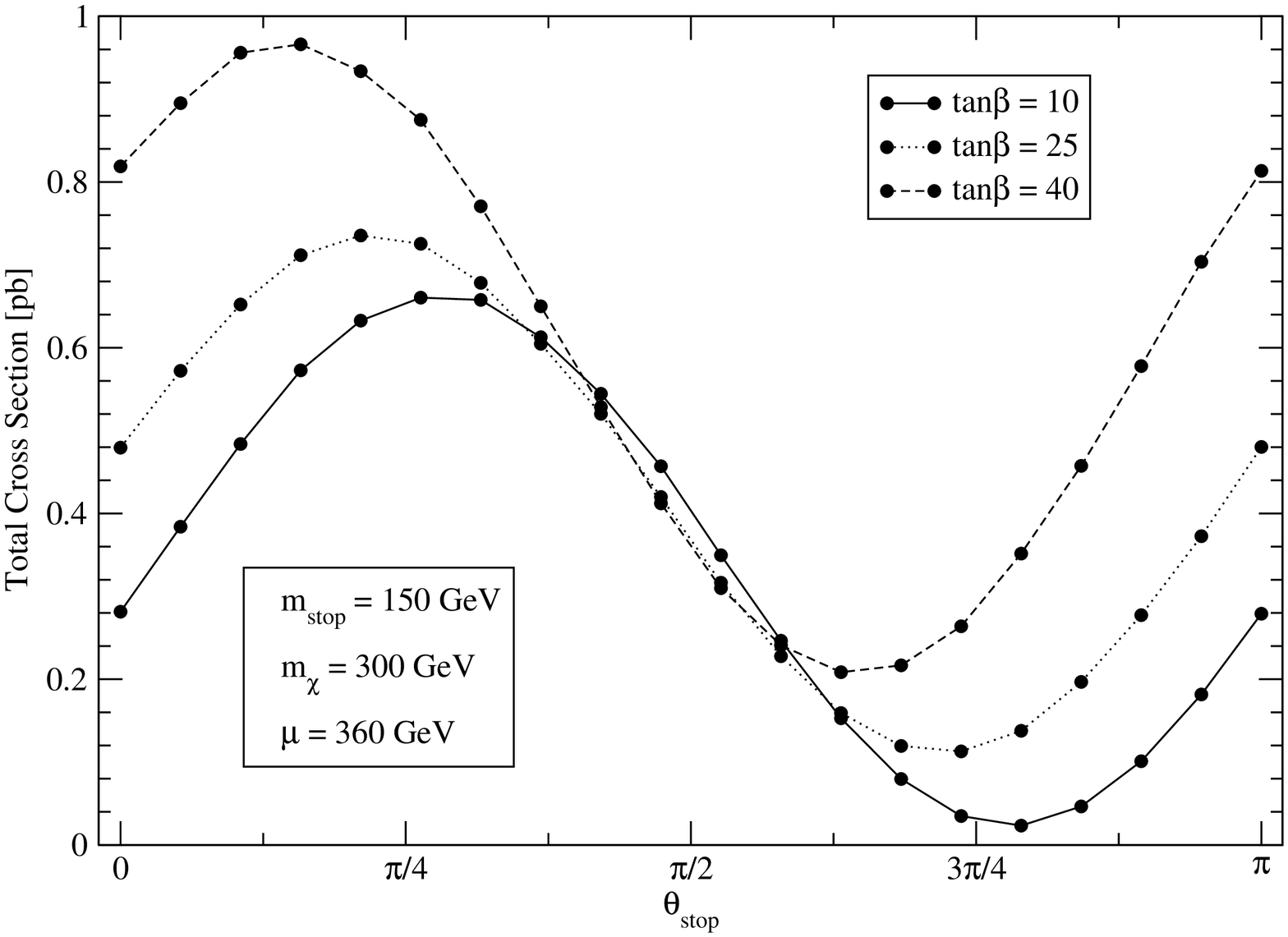, width=0.48\textwidth, angle=0}
  \label{fig:parameter:b}}\\
  \subfigure[]{
  \epsfig{file=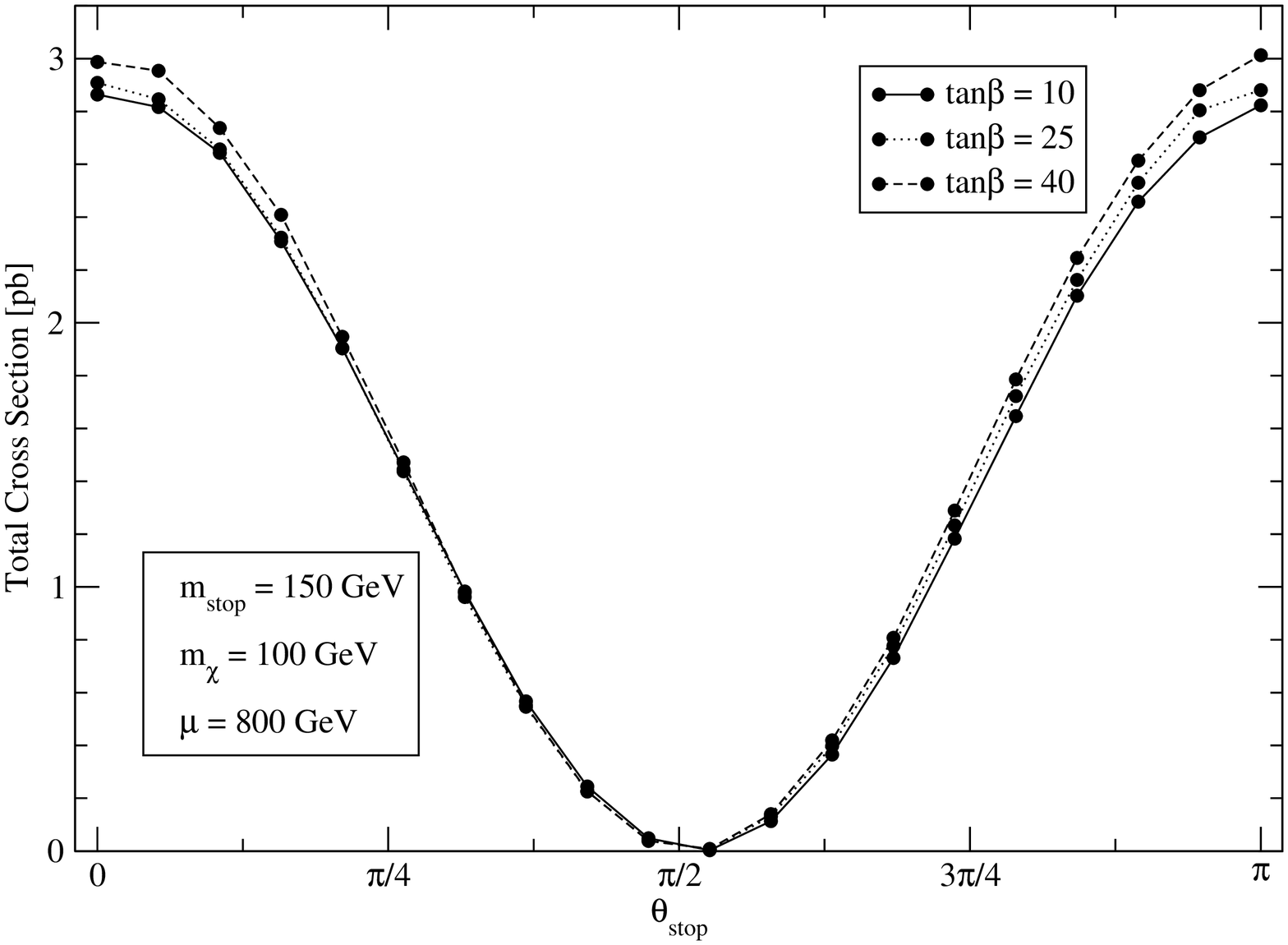, width=0.48\textwidth, angle=0}
  \label{fig:parameter:c}}\hfill
  \subfigure[]{
  \epsfig{file=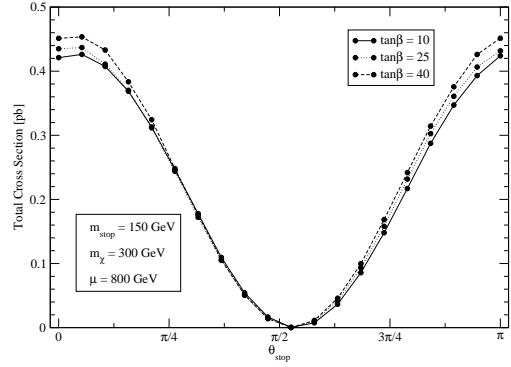, width=0.48\textwidth, angle=0}
  \label{fig:parameter:d}}
\end{minipage}\hfill
\caption{Parameter dependence of the total cross section at Born level}
\label{fig:parameter}
\end{figure}

Remaining at Born level it is already possible to extract
relevant e.w. information from the process.
Although some preliminary search of this kind already exists \cite{Beccaria:2006wz,Bozzi:2007me},
we will devote this Section to a brief updated summary of the main information that could
be derived from this approximate treatment.\newline
Indeed, at Born level a limited set of parameters affects the determination of physical observables.
Besides the values of the stop and chargino masses, which are obviously crucial for
the definition of the production threshold, other SUSY parameters contribute 
to the the coupling $b\tilde t_a^{~}\chi_i^-$, Eq.(\ref{couplings}). 
While $\tan\beta$ explicitly appears in the various terms, the chargino 
mixing matrices $Z_{ij}^\pm$ depend in a non-trivial way on parameters of 
the chargino mass matrix:
\begin{eqnarray}
  \label{chargino_mass_matrix}
  X=\left( \begin{array}{cc} M_2 & \sqrt{2}M_W\sin\beta \\ \sqrt{2}M_W\cos\beta & \mu \end{array} \right)
\end{eqnarray}
Moreover, for production of physical stops, the mixing angle $\theta_{\tilde
  t}$ mixes the various terms of Eq.(\ref{couplings}). In conclusion, it is 
possible to identify a set of independent parameters which determine the amplitude at Born level:
\begin{equation}
  \tan\beta \qquad M_2 \qquad \mu \qquad m_{\tilde t} \qquad \theta_{\tilde t}
\end{equation}
The chargino masses are determined by a combination of $M_2$, $\mu$ 
and $\tan\beta$. Since to perform a parameter analysis of the process it 
seems reasonable to fix all the masses, it is possible to trade, e.g., $M_2$ 
for $m_{\chi_1}$ and $\mu$ for $m_{\chi_2}$, depending on the chargino of 
the final state. \newline
Given these premises, the process of production of the lightest stop and 
chargino $bg \to \tilde t_1 \chi_1$ will now be analyzed at Born level to
investigate possible dependences on supersymmetric parameters. Rather than 
looking for a dependence of the cross section on the stop or chargino masses 
that we assumed to be experimentally known from previous discovery, we looked 
for dependences on $\tan\beta$, $\mu$ and $\theta_{\tilde t}$. The results of 
the analysis are shown in Fig.~\ref{fig:parameter}: All panels show the 
dependence of the total cross section on the mixing angle $\theta_{\tilde t}$ 
for different values of $\tan\beta$, at different values of $\mu$. In
particular, Figs.~\ref{fig:parameter:a} and \ref{fig:parameter:b} show the 
results obtained for low values of $\mu$, close to the chargino mass, while in 
Figs.~\ref{fig:parameter:c} and \ref{fig:parameter:d} $\mu$ has been pushed to 
the value of 800Gev, which is high compared to $m_{\chi_1}$. It is possible to 
notice that the cross section depends very strongly on the value of
$\theta_{\tilde t}$ and that there is always a value of the angle for which
the cross section drops near to zero. In Fig.~\ref{fig:parameter:a}, where 
the low threshold allows a cross section of the order of the pb, it is
possible to see that $\sigma$ changes from $\sim$6 pb for $\theta_{\tilde t}
\simeq \pi/8$ to less than 0.5 pb for $\theta_{\tilde t} \simeq 5\pi/8$ 
(when $\tan\beta = 40$). Therefore there are regions of the parameter space
where, even if the masses of final state particles are very low, the stop
mixing angle pushes the cross section to nearly undetectable levels. 
The dependence on $\theta_{\tilde t}$ can indeed be understood looking 
at the amplitude, which is a sum of terms of the form:
\begin{equation}
  A_1 \cos\theta_{\tilde t} + A_2 \sin\theta_{\tilde t}
\end{equation}
Depending on the values of $A_1$ and $A_2$, the squared amplitude 
generates the curves shown in Fig.~\ref{fig:parameter}.\\
The results also depend in a weaker and less trivial way on $\tan\beta$. 
The cross section is pushed to somewhat higher values as $\tan\beta$
increases, but to different extents in the various considered cases:
the variation of the total cross sections with different values of $\tan\beta$ 
are strongly affected by the choice of $\mu$ resulting in a very mild 
dependence on $\tan\beta$ for high values of $\mu$ and viceversa,
so that the determination of this supersymmetric parameter from this process 
could be ambiguous, unless the rate turns out to be  larger than a certain 
``threshold'' value. Further constraints coming from other processes could 
however limit the range of $\tan\beta$, and the determination of the two 
remaining parameters through the analysis of this process would then be relevant.

\subsection{One-Loop Amplitude}

\begin{figure}[t]
\centering
\epsfig{file=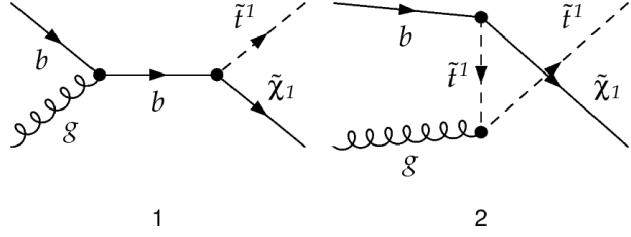, width=0.5\textwidth, angle=0}
\caption{Born Diagrams: s-channel bottom exchange and u-channel stop exchange}
\label{fig:bgchistop_tree}
\end{figure}

\begin{figure}
\centering
\epsfig{file=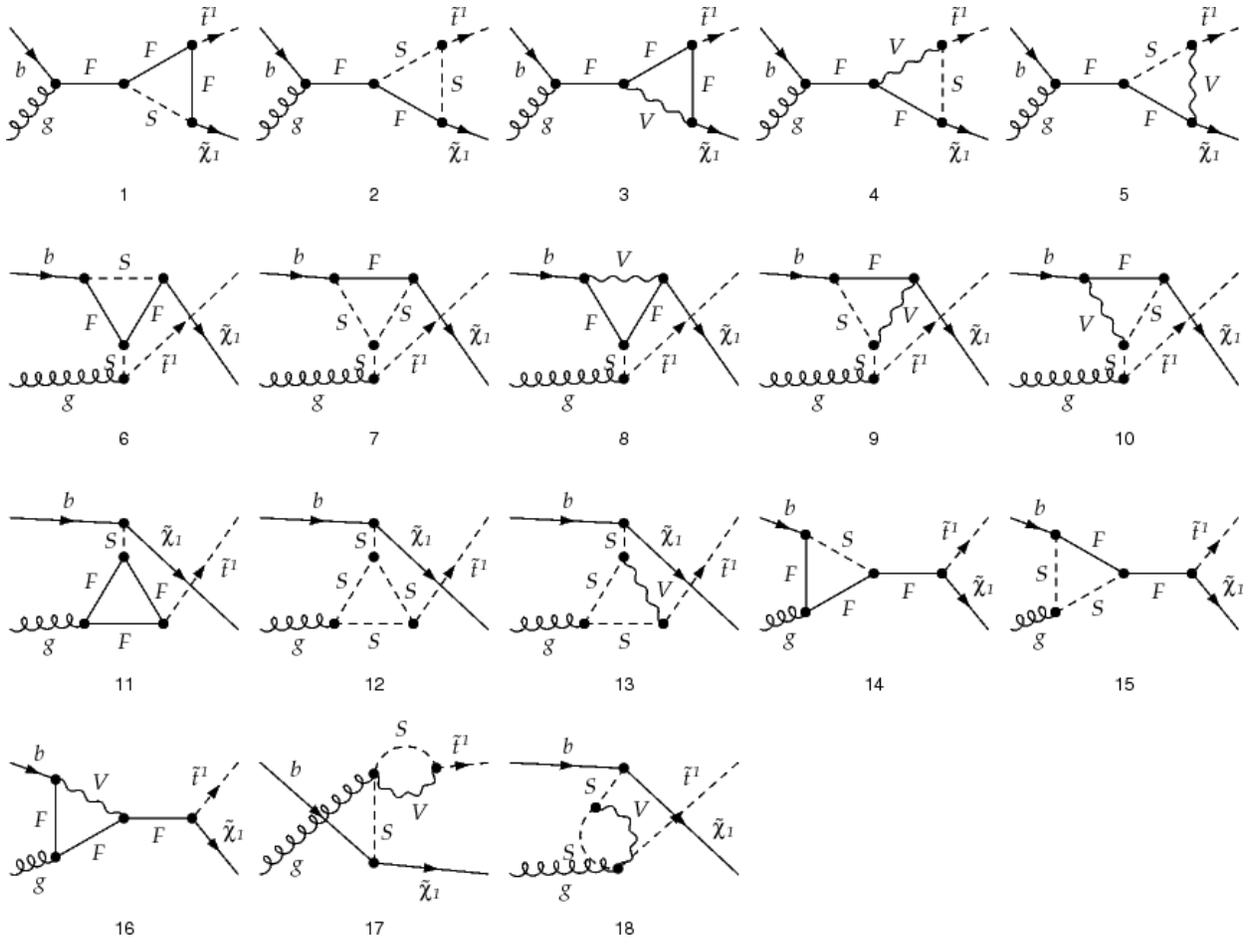, width=\textwidth, angle=0}
\caption{Triangle and 4-leg diagrams}
\label{fig:bgchistop_triangle}
\end{figure}

\begin{figure}
\centering
\epsfig{file=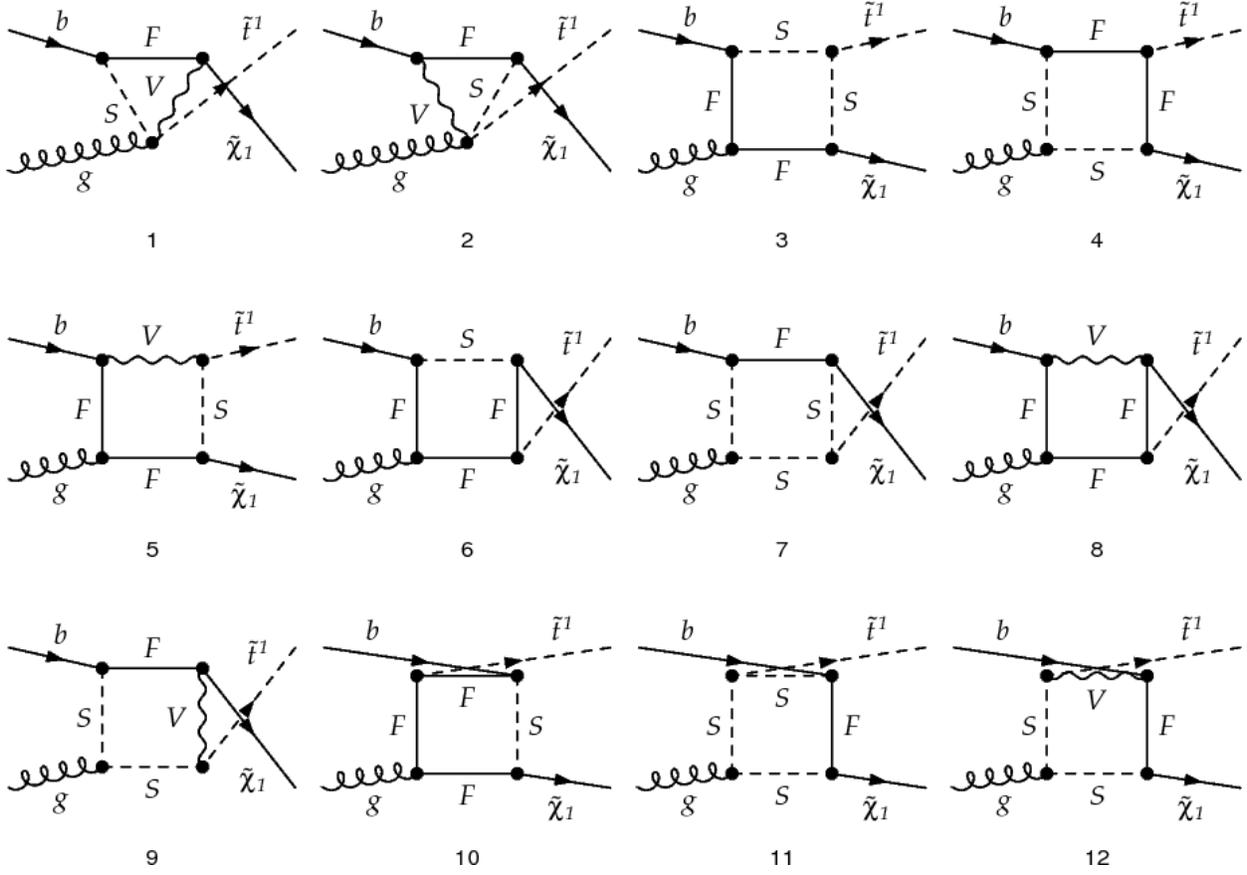, width=\textwidth, angle=0}
\caption{Triangle and box diagrams}
\label{fig:bgchistop_box}
\end{figure}

For the calculation of the one-loop amplitude we use the on-shell scheme;
the one-loop electroweak terms can be classified in:

\begin{itemize}
  \item[---] counter terms for $b,\tilde{t}_a,\chi^-_i$ lines,
  coupling constants and mixing elements, all of them
  being expressed in terms of self-energy diagrams;
  \item[---] self-energy corrections for $b$ and $\tilde{t}_a$
  propagators;
  \item[---] s-channel left and right triangles;
  \item[---] u-channels bubbles with 4-leg couplings and up, down
  triangles;
  \item[---] direct boxes, crossed boxes, twisted boxes;
\end{itemize}
and the related diagrams are shown in Figs.~\ref{fig:bgchistop_tree}-\ref{fig:bgchistop_box}.

Since the complete expressions of the various conter-terms and self
energies are rather involved we list them separately in the App.~\ref{AppendixA}.
All the contributions of the counter-terms and self energies, together with
the virtual vertex and box diagrams have been computed using the
usual decomposition in  terms of Passarino-Veltman functions and 
the complete amplitude has been implemented in the numerical code \verb+TigreMC+.\\

We have checked the cancellation of the UV divergences
among counter terms, self-energies and triangles,
this cancellation occuring
separately for s-channel and for u-channel,
as well as for gauge-left, gauge-right, Yukawa-left and
Yukawa-right sectors separately.\\

Another useful check can be done using the high energy behaviour of the amplitudes.\\
High energy rules \cite{Beccaria:2002cf} predict the logarithmic
behaviour of these amplitudes at one-loop level.
They use splitting functions for external particles
$b$, $\tilde{t}_{L,R}$, $\chi^-_i$
and Renormalization Group effects on the parameters
appearing in the Born terms. They read:
\bqa
F_{-++}(\tilde t_L^{~})&=&-g_s\sqrt{2}
A^L_i(\tilde t_L^{~})\sin{\theta\over2}
\{1+{\alpha\over4\pi}\{\nonumber\\
&&
{1+26c^2_W\over18s^2_Wc^2_W}~\log\frac{s}{M_W^2}
-[{m^2_t\over2s^2_WM^2_W}(1+\cot^2\beta)+
{m^2_b\over2s^2_WM^2_W}(1+\tan^2\beta)]\log\frac{s}{M_W^2}\nonumber\\
&&
-\{{1\over2s^2_W}\log^2{-u\over m^2_W}+\log^2{-u\over m^2_Z}]
+{1-10c^2_W\over36s^2_Wc^2_W}~\log^2{-t\over m^2_Z}~
~\}\}
\label{fang0}
\eqa
\bqa
F_{-++}(\tilde t_R^{~})&=&-~g_s\sqrt{2}
A^L_i(\tilde t_R^{~})\sin{\theta\over2}
\{1+{\alpha\over4\pi}.\nonumber\\
&&\{
-[{1\over3c^2_W}]\log^2\frac{s}{m^2_Z}
-[{1\over9c^2_W}]\log^2\frac{-t}{m^2_W}
\nonumber\\
&&
+~{1-4c^2_W\over12s^2_W c^2_W}[\log^2\frac{-u}{M_Z^2}]
-~~{1\over2s^2_W}[\log^2\frac{-u}{M_W^2}]\}~\}
\label{fang1}
\eqa
\bqa
F_{+--}(\tilde t_L^{~})&=&g_s\sqrt{2}
A^R_i(\tilde t_L^{~})\sin{\theta\over2}
\{1+{\alpha\over4\pi}.\nonumber\\
&&\{
-[{1+2c^2_W\over12s^2_Wc^2_W}]\log^2\frac{s}{m^2_Z}
-[{1\over2s^2_W}]\log^2\frac{s}{m^2_W}
\nonumber\\
&&+{1\over18c^2_W}[\log^2\frac{-t}{M_W^2}]
-~{1\over6s^2_W}[\log^2\frac{-u}{M_W^2}]
~\}~\}
\label{fang2}
\eqa
The logarithmic part of the gaugino amplitude (\ref{fang0}) is similar to the
one obtained for the process $bg\to tW^-$ with transverse $W$ \cite{Beccaria:2006dt}.
Higgsino amplitudes $F_{-++}(\tilde t_R^{~})$, $F_{+--}(\tilde t_L^{~})$
in (\ref{fang1},\ref{fang2}) get logarithmic terms similar to the ones in both $bg\to tW^-$
for longitudinal $W$ and $bg\to tH^-$ \cite{Beccaria:2004xk}. One notices that there is no linear
logarithmic contribution , but only quadratic logarithmic terms, in these Higgs or
Higgsino type of amplitudes. The coefficients of these quadratic logarithms
are of pure gauge origin and do not involve any free parameter.\\

Taking our complete one-loop computation and retaining only the logarithmic
parts of the B,C,D Passarino-Veltman functions appearing in the various diagrams, we do
recover the above expressions for the 3 types of leading amplitudes.\\

We now give illustrations of the various features mentioned above
for the process $bg\to\tilde{t_1}\chi_1$ with production of the
lightest stop and chargino. We choose two typical benchmark MSSM points 
called LS1 and LS2 whose characteristics are shown in Tab.{\ref{tab:bench}}
together with those of all the benchmark points that we have used for the analysis (see next Section for more details).\\

\begin{figure}
\centering
\epsfig{file=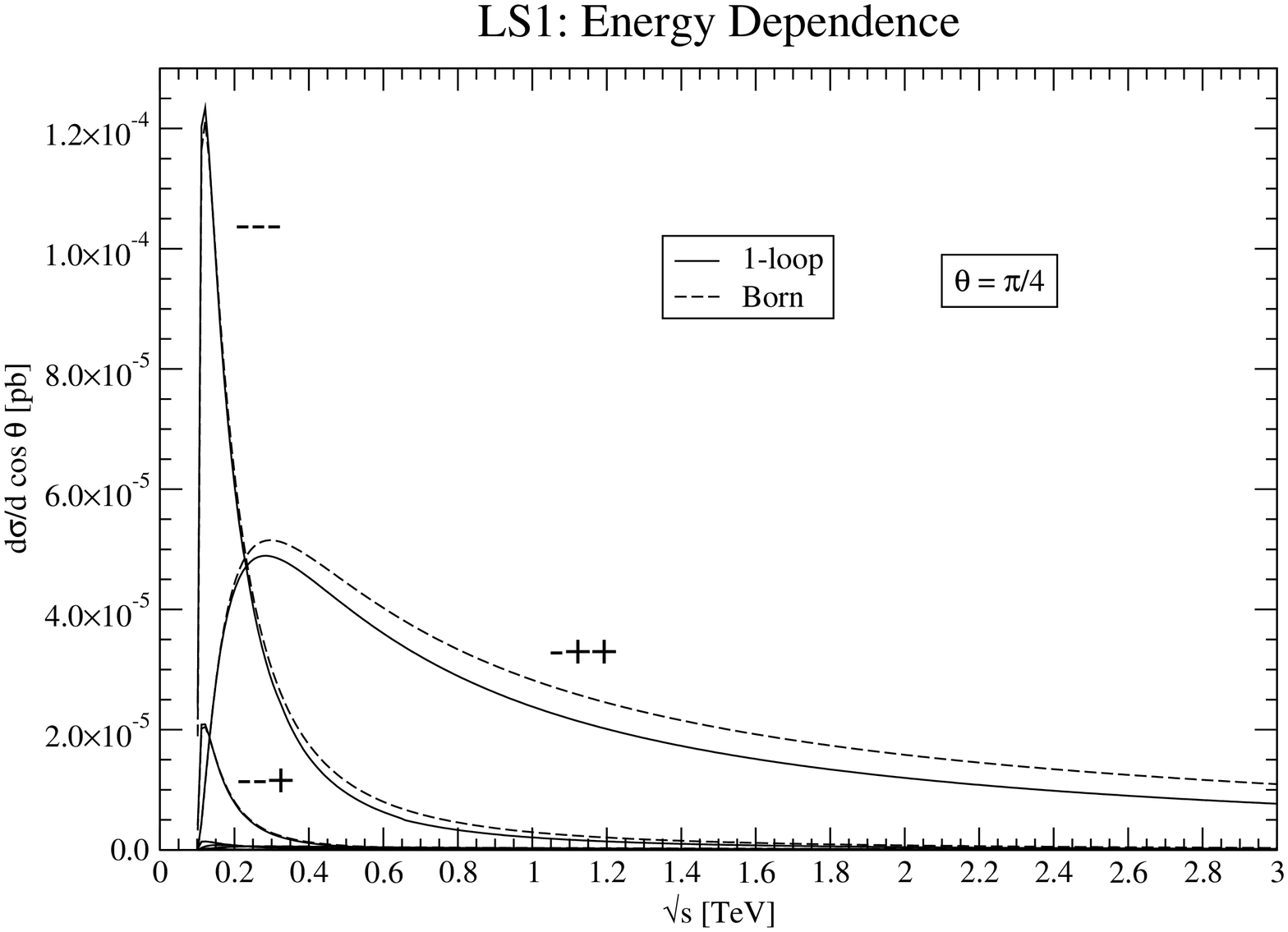, width=0.9\textwidth, angle=0}\\
\epsfig{file=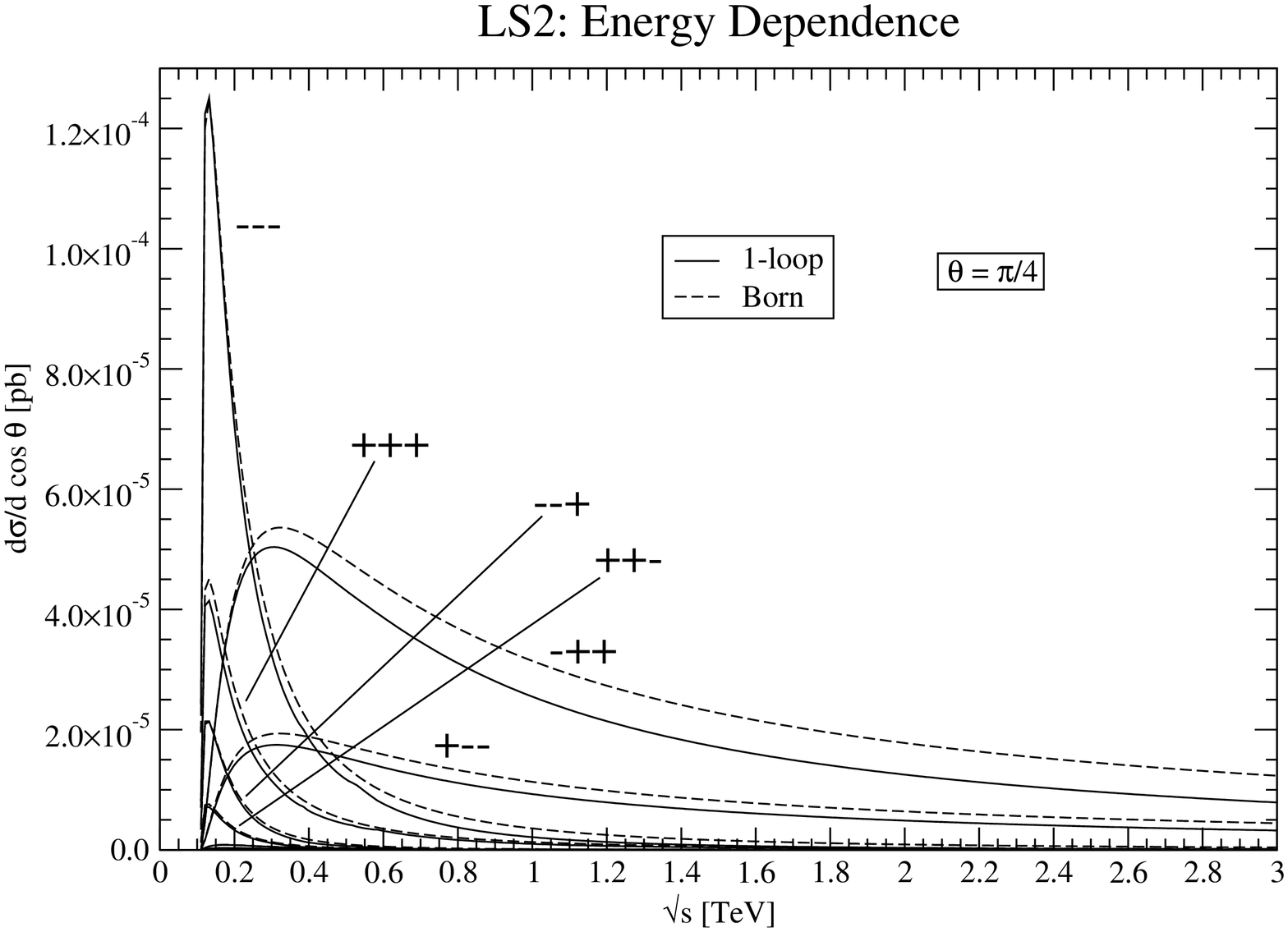, width=0.9\textwidth, angle=0}
\caption{Energy dependence for leading helicity amplitudes in LS1 and LS2.}
\label{fig:energy}
\end{figure}

\begin{figure}
\centering
\subfigure{
\begin{minipage}{0.5\textwidth}
\epsfig{file=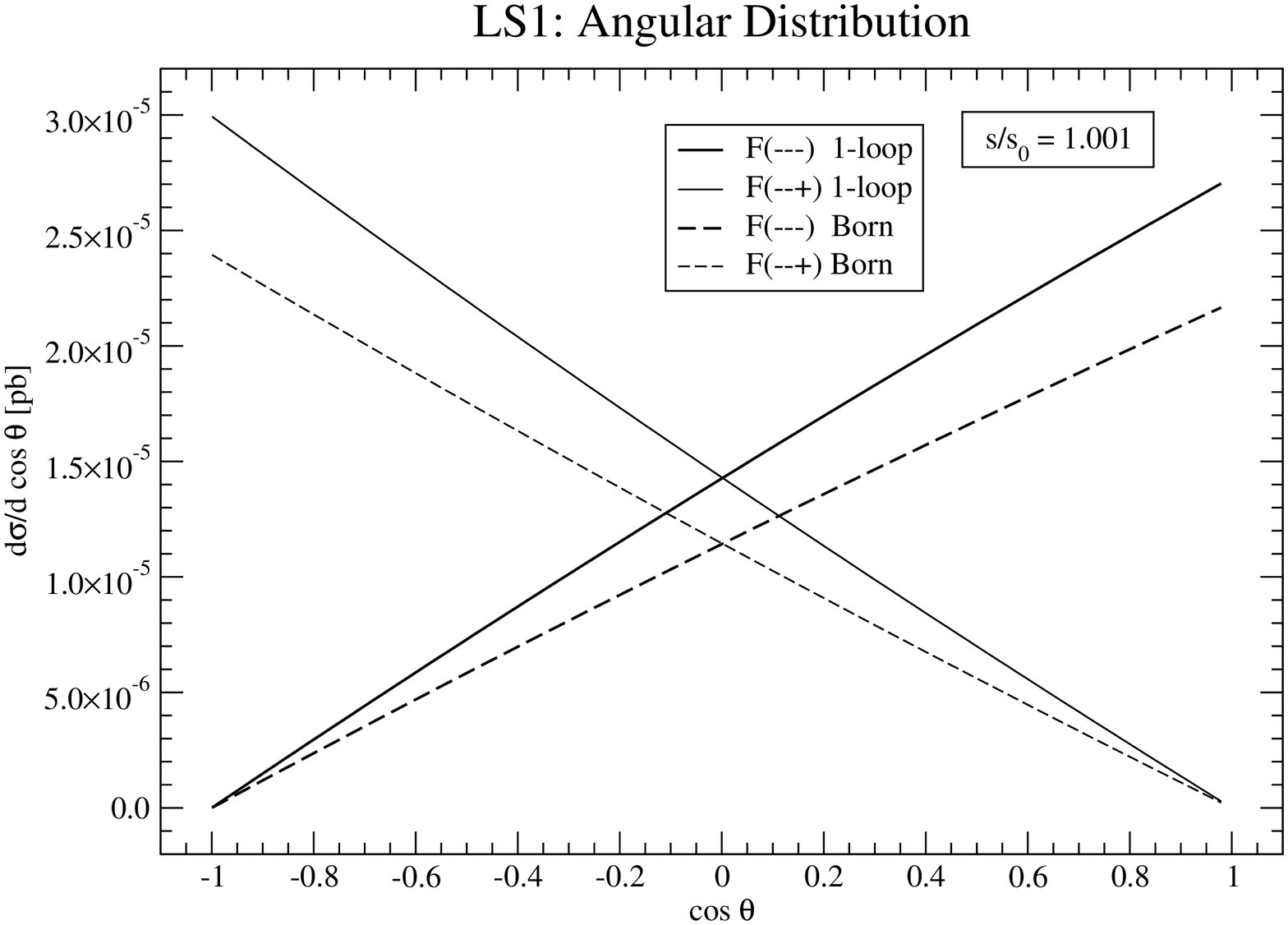, width=0.9\textwidth, angle=0}\\[5pt]
\epsfig{file=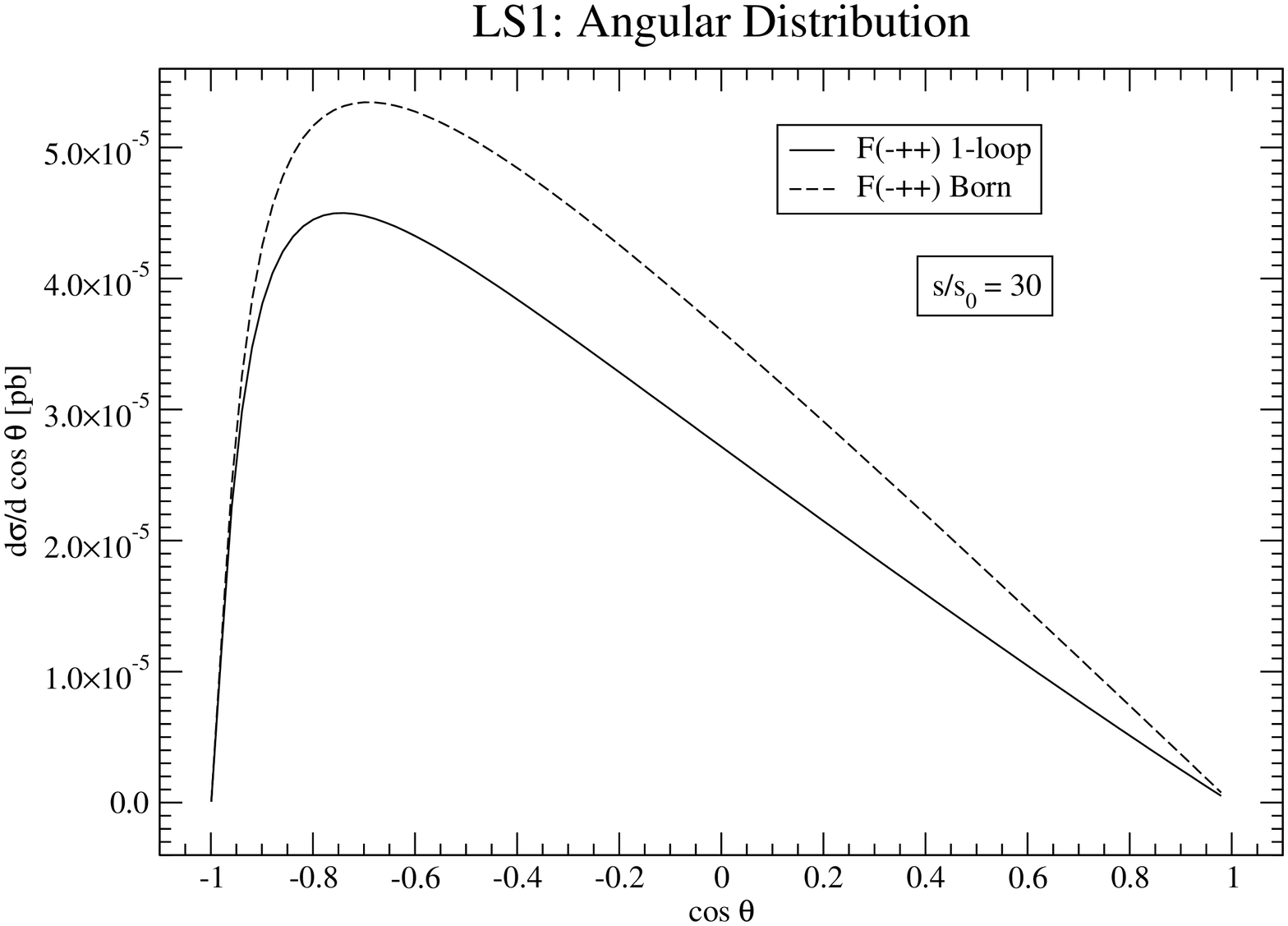, width=0.9\textwidth, angle=0}
\end{minipage}
\begin{minipage}{0.5\textwidth}
\epsfig{file=./Figures/LS2_angular_low.eps, width=0.9\textwidth, angle=0}\\[5pt]
\epsfig{file=./Figures/LS2_angular_high.eps, width=0.9\textwidth, angle=0}
\end{minipage}}
\caption{Angular distribution for leading helicity amplitudes in LS1 and LS2 in the low and high energy limits.}
\label{fig:angular}
\end{figure}

\begin{table}[htbp]
 \begin{tabular}{|c|ccccc||c|c|}
 \hline
 ~mSUGRA scenario~ & $m_0$ & $m_{1/2}$ & $A_0$ & $\tan\beta$ & $\textrm{sign } \mu$ & $\quad m_{\widetilde{t}_1} \quad$ & $\quad m_{\chi_1}\quad$  \\
 \hline
 LS1         & 300           & 150 & -500           & 10 & + & 214.6 & 103.6 \\
 LS2         & 300           & 150 & -500           & 50 & + & 224.7 & 106.9 \\
 SPS5        & 150           & 300 & -1000          & 5  & + & 279.0 & 226.2 \\
 SU1         & 70            & 350 & 0              & 10 & + & 566.4 & 255.7 \\
 SU6         & 320           & 375 & 0              & 50 & + & 634.1 & 279.7 \\
 \hline  
 \end{tabular}
\caption{mSUGRA benchmark points and masses of the lightest stop and chargino (all the values are in GeV)}
\label{tab:bench}
\end{table}
Fig.~\ref{fig:energy} shows the energy dependence of each helicity amplitude
from threshold to high energy for a given (central) angle $\theta=\pi/4$.
For each amplitude two curves represent the Born and the full one
loop result. One can check that they confirm the expectations
described in the previous subsection, namely the nature of:
\begin{itemize}
\item[---] the dominant amplitudes at low energy
\item[---] the dominant amplitudes at high energy
\end{itemize}
The size of the one-loop correction is of the order of few percent, and one
sees that the high energy behaviour is quickly reached as soon as the
threshold is crossed.\\
The difference between the LS1 and LS2 cases is due to the increase
in the final masses and in the change in the stop and in the chargino
mixings. In particular for LS2 the R chirality amplitudes are less
depressed because of the the difference in $\tan\beta$ and in the mixing
element $Z^-_{2i}$, which increase the value of $A^R_i(\tilde t_L^{~})$ (see Eq.(\ref{couplings})) and this can be clearly seen both
at low and at high energies.\\
Fig.~\ref{fig:angular} gives the angular distributions at low energy ($s/s_{\rm thresh.}=1.001$) and at high
energy ($s/s_{\rm thresh.}=30$). At low energy the leading amplitudes give indeed the 
expected $(1+\cos\theta)$ and $(1-\cos\theta)$ distributions,
whereas at high energy one tends to a limiting $(1-\cos\theta)$
distribution, at least away from purely backward scattering. 
The one-loop corrections make only little changes
in the shape of the angular distributions as expected from the
Sudakov rules.\\

\subsection{QED radiation}
\label{subsection:qed}

The ${\cal O}(\alpha)$ electroweak corrections include contributions 
from virtual and from real photon emission. The virtual photon exchange 
diagrams belong to the complete set of electroweak virtual corrections, 
and are necessary for the gauge invariance of the final result. 
The singularities associated with the massless nature of the photon have 
been regularized by introducing a small photon mass $m_\gamma$. 
The real radiation contribution has been split into a soft part, 
derived within the eikonal approximation, where the photon energy has 
been integrated from the lower bound $m_\gamma$ to a maximum cut-off
$\Delta E$, and into a hard part, integrated from
the minimum photon energy $\Delta E$ to the maximum 
allowed kinematical value. 
The soft real contribution 
contains explicitly the photon mass parameter $m_\gamma$ while the
hard part can be  calculated with a massless external photon. The
complete  
matrix element for real radiation, including fermion
mass effects,
has been calculated analytically with the help 
of \verb+FeynArts+~\cite{Kublbeck:1990xc}
and  \verb+FormCalc+~\cite{Hahn:1998yk}.

The logarithmic terms containing $m_\gamma$ cancel 
exactly in the sum of virtual and soft real part, leaving only
polynomial spurious terms, which approach zero at least as
$m_\gamma^2$. We have numerically
checked the cancellation by taking the limit $m_\gamma \to 0$ of our computation.
The large collinear logarithms containing the bottom mass
are only partially cancelled when real and virtual corrections are summed
together, but they can be absorbed into the
definition of the parton distribution functions (PDFs). This can be
achieved redefining the bottom PDF according to a factorization
scheme. In the $\overline{\mbox{MS}}$ (DIS) scheme such redefinition reads~\cite{Baur:1998kt} 
\begin{eqnarray}
b(x, \mu) &\to& b(x,\mu) \left \{ 1 - \frac{\alpha}{\pi} e^2_b\left [ 1 -\ln
  \delta_s - \ln \delta^2_s + \left ( \ln \delta_s + \frac{3}{4}
  \right ) \ln \left ( \frac{\mu^2}{m_b^2}  \right ) - \frac{1}{4}
    \lambda_{\mbox{\tiny FC}} f_1  \right]  \right \} \nonumber \\
&-& \frac{\alpha}{2 \pi} e^2_b\int_{x}^{1-\delta_s}\; \frac{dz}{z}  \;
b\left( \frac{x}{z}, \mu \right ) \left [ \frac{1+z^2}{1-z} \ln \left
  (\frac{\mu^2}{m^2_b} \frac{1}{(1-z)^2} \right ) - \frac{1+z^2}{1-z}
+ \lambda_{\mbox{\tiny FC}} f_2\right ],
\end{eqnarray}
with $\lambda_{\mbox{\tiny FC}} = 0$ ($\lambda_{\mbox{\tiny
    FC}}=1$). $\mu$ is the factorization scale, $\delta_s = 2 \Delta
E / \sqrt{s}$, while $e_b$ is the bottom charge.  $f_1$ and $f_2$
are defined as follows,
\begin{eqnarray}
f_1 &=& 9 +\frac{2}{3} \pi^2 +3 \ln \delta_s -2 \ln^2 \delta_s,
\nonumber \\
f_2 &=& \frac{1+z^2}{1-z} \ln \left (\frac{1-z}{z} \right )
-\frac{3}{2} \frac{1}{1-z} +2 z +3.
\end{eqnarray}  
The calculation
of the full $\mathcal{O}(\alpha)$ corrections to any hadronic observable must include  
QED effects in the DGLAP evolution equations. Such effects are taken
into account in the MRST2004QED PDF~\cite{Martin:2004dh}.
This set is however NLO QCD, while our computation is leading order QCD. Therefore,
analogously to~\cite{WjetProd}, a LO QCD PDF set has been chosen, namely the
CTEQ6L~\cite{Pumplin:2002vw}. This choice is justified by the fact that QED effects
are known to be small~\cite{Roth:2004ti}.
In the numerical analyses
we have used the $\overline{\mbox{MS}}$
factorization scheme at the 
scale $\mu = (m_{\tilde{t}_1}+m_{\tilde{\chi}^-_1})$.  It is worth to
mention that the dependence
of the full $\mathcal{O}(\alpha)$
contribution on  the factorization scheme is rather weak. 
Indeed, if the DIS factorization
scheme is used instead of $\overline{\mbox{MS}}$,
the differences in the numerical value of the one-loop electroweak effects
are of the order of 0.01\% in all the considered mSUGRA benchmark points.

\begin{figure}
\centering
\epsfig{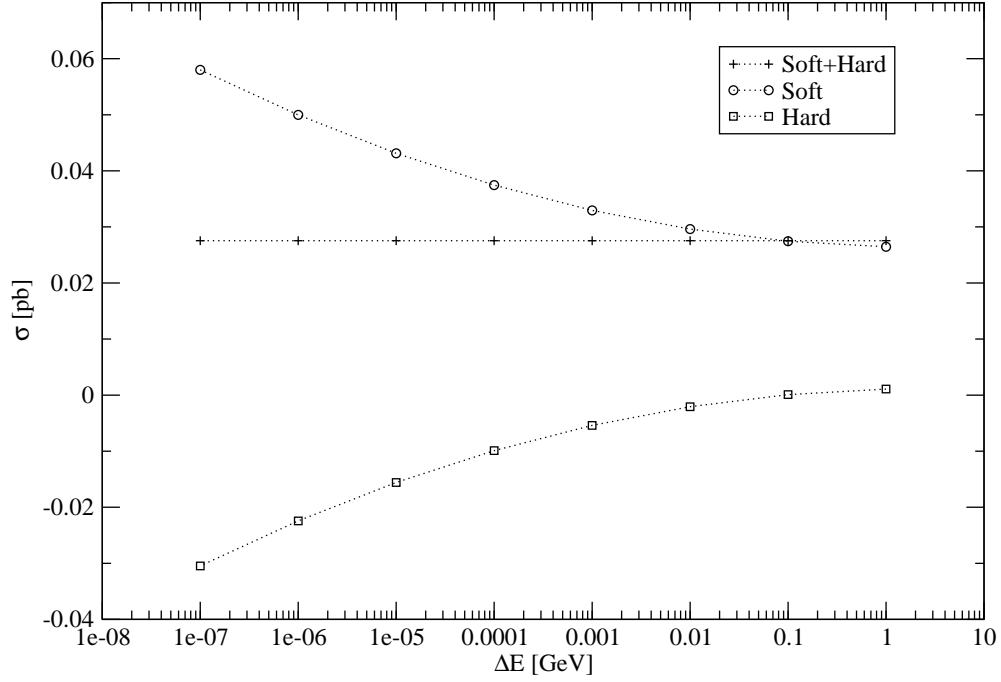}\\[50pt]
\epsfig{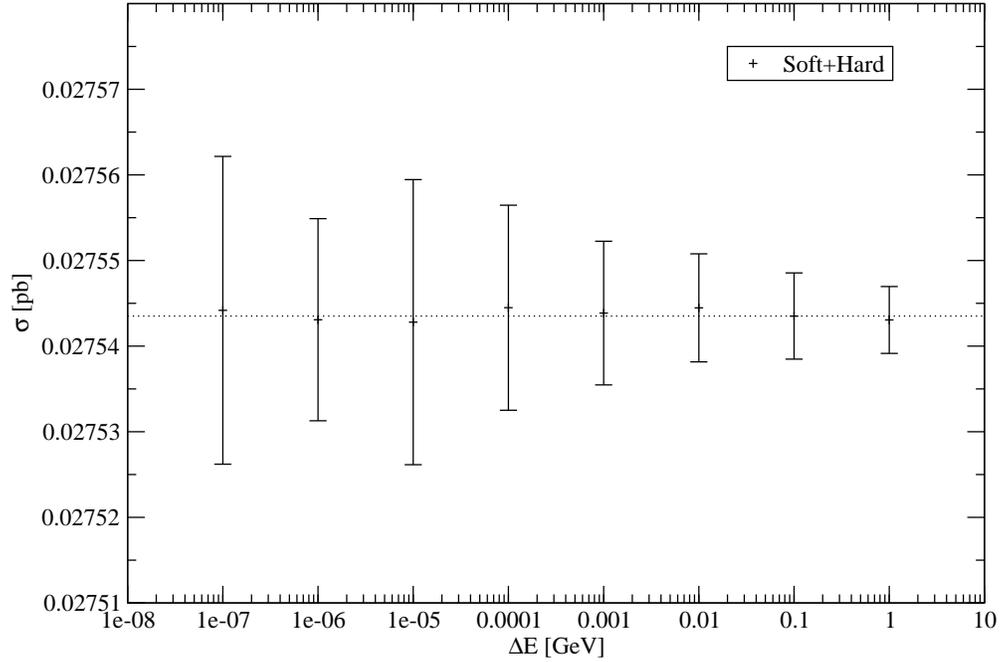}
\caption{Upper panel: dependence of the ${\cal O}(\alpha)$ 
soft plus virtual and hard cross sections on the soft-hard 
separator $\Delta E$. Lower panel: independence of the sum of ${\cal O}(\alpha)$ soft plus 
virtual and hard cross sections 
of the separator $\Delta E$.}
\label{fig:QEDcheck}
\end{figure}

The final cross section has to be independent of the 
fictitious separator $\Delta E$, for sufficiently small $\Delta E$ values. 
This has been checked numerically to hold for $\Delta E \leq 1$~GeV, as shown in Figure~\ref{fig:QEDcheck} 
(lower panel), 
despite the strong sensitivity to $\Delta E$ of the soft 
plus virtual and of the hard 
cross section separately, as shown in Figure~\ref{fig:QEDcheck} (upper panel).

Similarly to what has been obtained in our previous works \cite{Beccaria:2007tc} and \cite{Beccaria:2008av},
QED contributions to the total cross section are positive with a relative size of the order of a few percent.

\section{One-Loop Results}
\label{sec:oneloopresults}

The distribution of the invariant mass of the final states $d\sigma/dM_{inv}$
has been evaluated at the one-loop electroweak level for a number of SUSY
benchmark points (assuming a mSUGRA supersymmetry breaking) with a wide 
variation of mass spectra. The obtained cross sections 
at the Born and one-loop level for five representative points (the "Light SUSY"
LS1, LS2, discussed in \cite{Beccaria:2006ir}, the ATLAS SU1 and SU6 \cite{DC2} and the
SPS5 "Light Stop scenario" \cite{SPS}) are collected in Tab.~\ref{tab:benchmarks_Xsec}: for
the realistic case of production of the lightest stop and chargino states
$\tilde t_1^{~}$ and $\chi_1^-$,
only the couple LS1 - LS2
give a cross section of order of the $pb$
(considering a global factor $2$, arising from the conjugate process),
that we shall consider in this paper as a reasonable limit for realistic detections at the LHC.
All the other input sets, including the SPS5 "Light Stop", give smaller
rates, and will not be further considered in what follows.\\
  
\begin{figure}
\centering
\epsfig{file=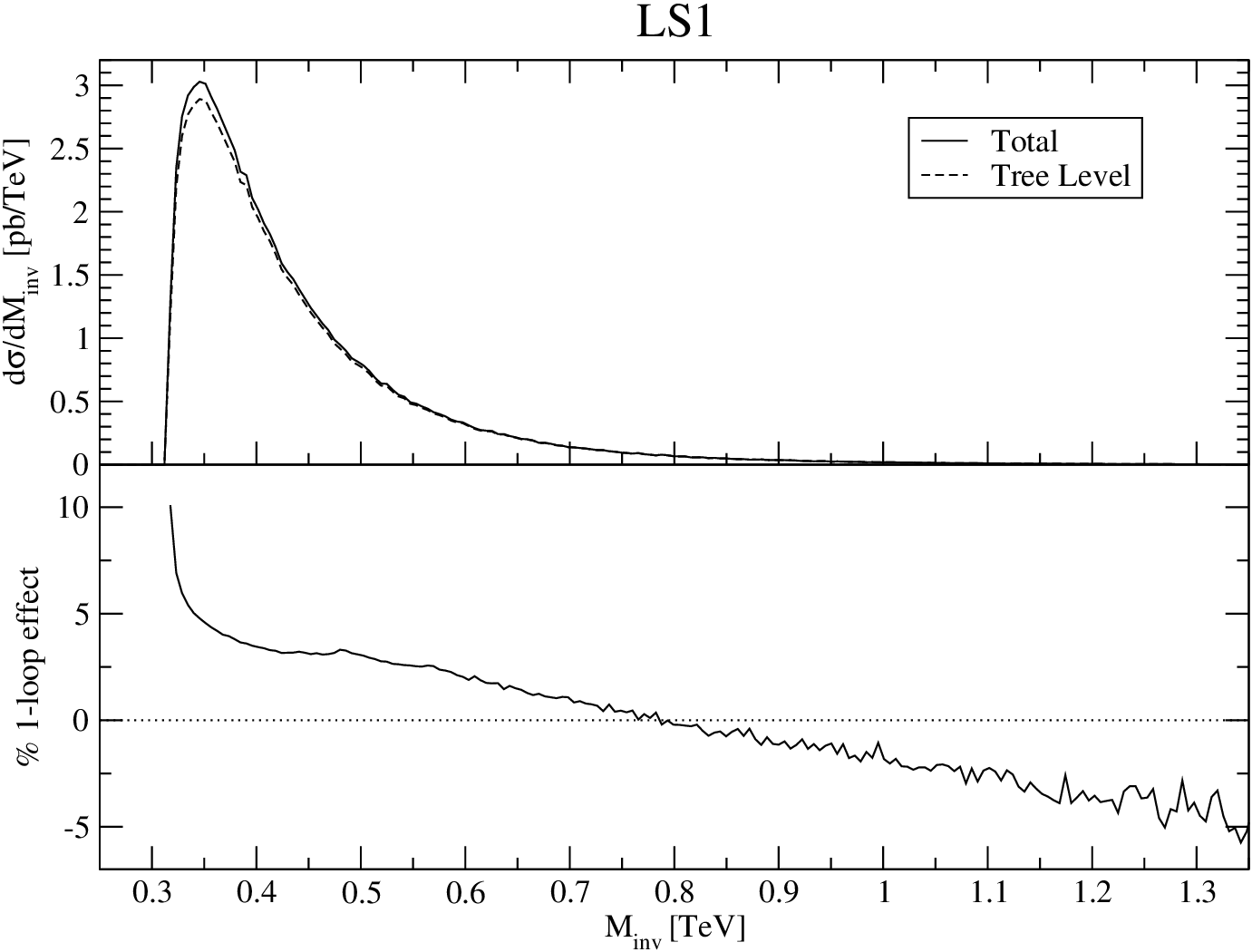, width=\textwidth, angle=0}
\caption{Differential distribution (upper panel) and percentage one-loop effect
in the LS1 point; $m_{\tilde t_1^{~}}= 214.5~GeV, m_{\chi_1}= 103.6~GeV $.}
\label{fig:1loop_LS1}
\end{figure}

\begin{figure}
\centering
\epsfig{file=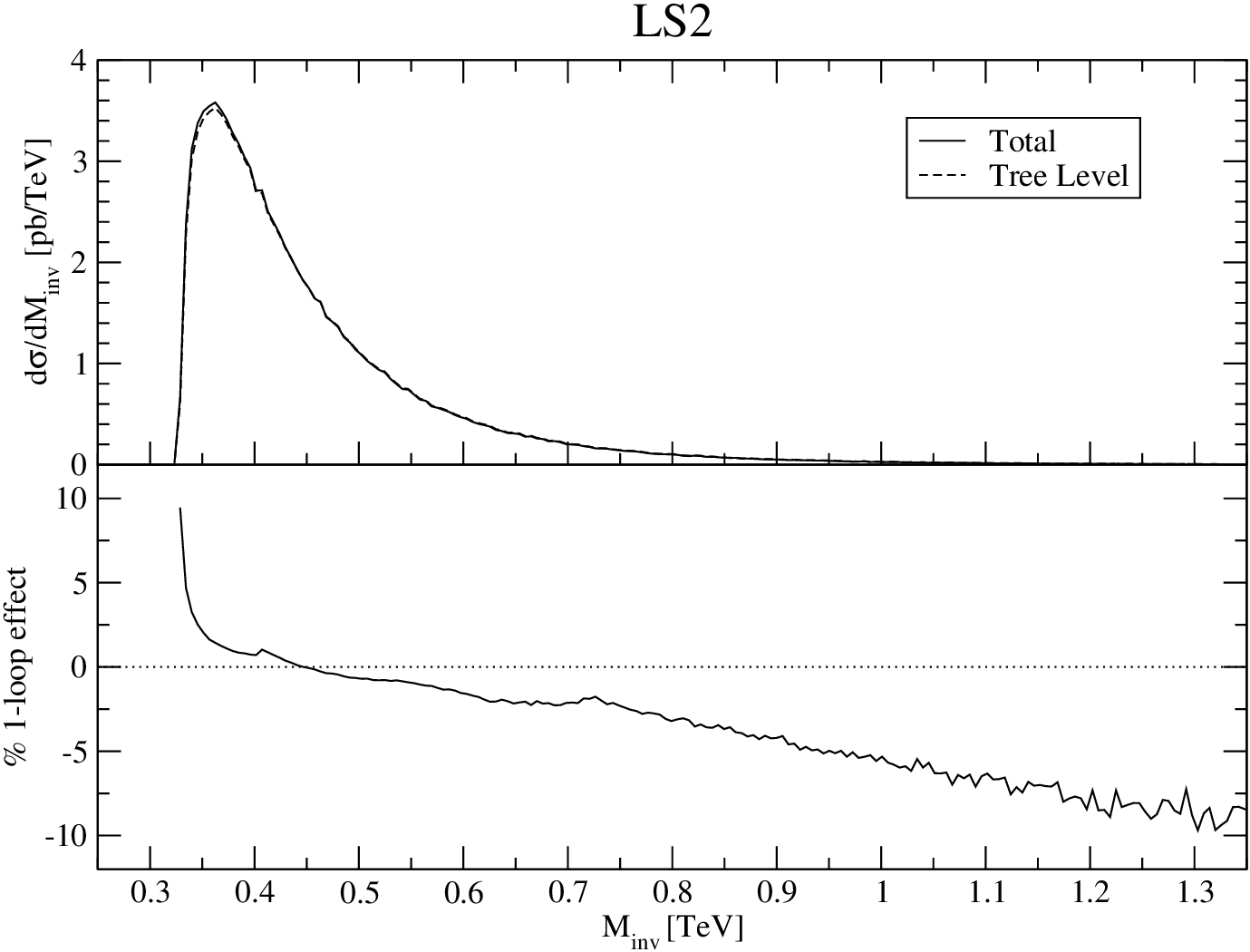, width=\textwidth, angle=0}
\caption{ Differential distribution (upper panel) and percentage one-loop effect
in the LS2 point; $m_{\tilde t_1^{~}}= 224.6~GeV, m_{\chi_1}= 106.9~GeV $.}
\label{fig:1loop_LS2}
\end{figure}

For what concerns the one-loop electroweak
corrections, we have found that they are generally small, of the
order of a relative few percent for all the considered scenarios.
\begin{table}[htbp]
\begin{tabular}{@{\extracolsep{5pt}}|c|ccc|}
\hline
mSUGRA scenario & $\sigma_{Born}$ & $\sigma_{1-loop}$ & ~$\%$ Effect~  \\
\hline
LS1  & 0.4287   & 0.4442   &  3.6 \\
LS2  & 0.5419   & 0.5436   &  0.3 \\
SPS5 & 0.05704  & 0.05810  &  1.8 \\
SU1  & 0.004052 & 0.004041 & -0.3 \\
SU6  & 0.002541 & 0.002576 &  1.4 \\
\hline
\end{tabular}
\caption{Total cross section at Born and loop level for the five considered benchmark points}
\label{tab:benchmarks_Xsec}
\end{table}
As an example of this behaviour in Figs.~\ref{fig:1loop_LS1} and \ref{fig:1loop_LS2}
we plot the differential distibutions for the LS1 and LS2 benchmark points,
(the points with highest cross sections): as one can notice, the
one-loop effect is positive in the low energy region
(near the production threshold) and drops to negative values increasing the final
invariant mass. The global effect on the totally integrated cross section,
being the result of the sum of two opposite contributions, is positive ($ 3.6 \% $) in the LS1 case, slightly smaller and below the $ 1 \%$ in the LS2 case.\newline
The conclusion of our analysis is thus, for what concerns the possibility
that NLO electroweak effects might affect the stop-chargino production
process, essentially negative in the chosen theoretical scheme, given the
fact that a realistic experimental accuracy of the measurements of the
various rates should hardly be better than, say, ten percent or more
(\cite{Clément:1132787}). In this spirit, it appears that the
complete dependence on the SUSY parameters can be satisfactorily
provided by the simple Born expressions of the process discussed in the previous Section.\newline
This conclusion is valid in the chosen theoretical scheme,
and is based on the relative smallness of the one-loop electroweak effects.
Clearly, the same conclusions cannot be drawn at this point for possible
different supersymmetric schemes. As a personal feeling, it seems unlikely
to us that strong one-loop effects might there arise, simply given the
unavoidably large sizes of the virtually exchanged sparticles. However, if
LHC discovered supersymmetry and reached a suitable experimental
accuracy, an extended analysis of the process that we have considered
might become definitely requested.

\section{Conclusions}

 
In this paper we have calculated the complete electroweak one-loop expression
of the stop-chargino process in the MSSM assuming a mSUGRA symmetry breaking 
scheme, to evidentiate possible realistically ``visible'' effects.
In our calculations we have verified the
fulfillment of a number of theoretical requests, including the reproduction
of asymptotic Sudakov expansions. This, we believe, should make our analysis 
reliable. As a result of our calculation we have concluded that the complete 
one-loop electroweak effect is of the relative few percent size, that would
make it hardly visible in a realistic LHC situation.
Given this result, the relevant e.w. information can be extracted from the Born expression of the rate.
We have examined its possible dependence on those supersymmetric
parameters, on which it depends, that cannot be directly measured from direct 
production, i.e. on the parameters $\mu$, $\tan\beta$ and $\theta_{\tilde t}$. Assuming  a
previous measurement of the stop and chargino masses, we have verified that
the dependence of the rates on $\theta_{\tilde t}$ and $\mu$ might be rather strong in the
case of light final state masses, and would influence the dependence on
$\tan\beta$. This would indicate that, given a light stop and chargino masses
picture, a measurement of the light stop-chargino process might provide an
original and useful type of constraints on the size of the relevant MSSM parameters.\newline

\appendix
\section{Counter Terms}
\label{AppendixA}

\begin{figure}[b]
\centering
\epsfig{file=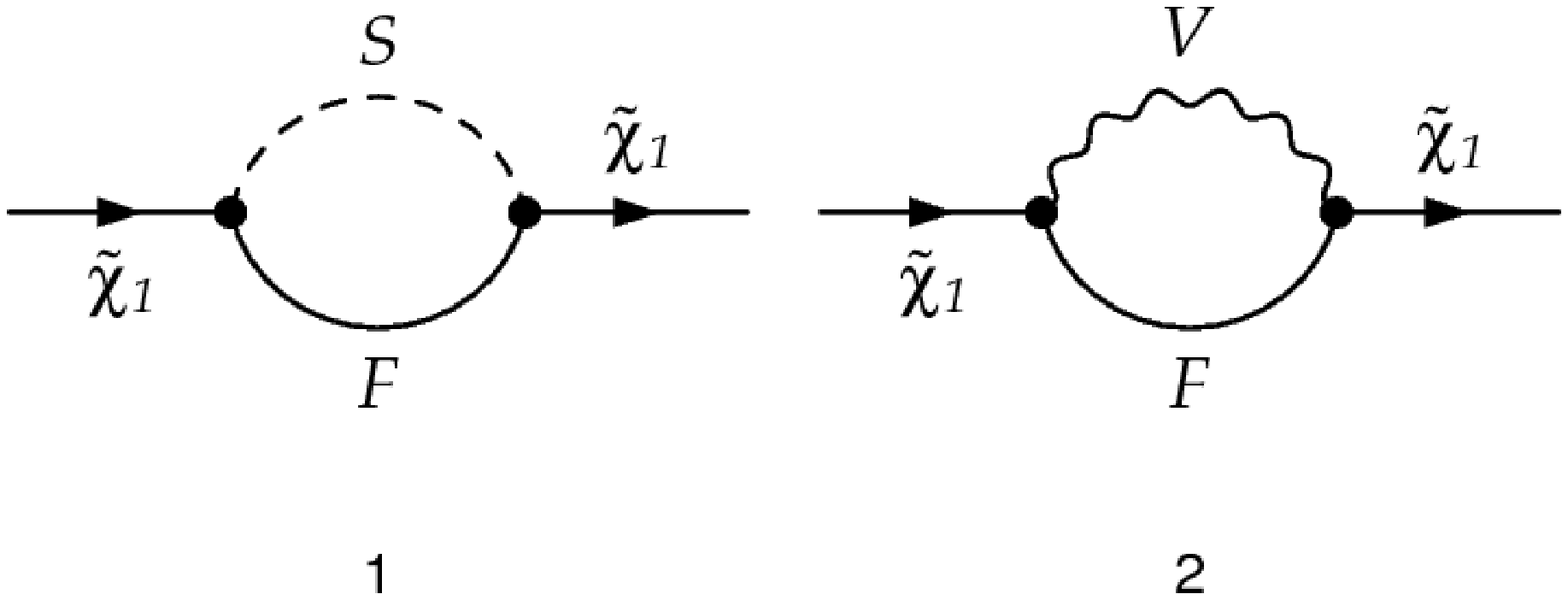, width=0.4\textwidth, angle=0}\hfill
\epsfig{file=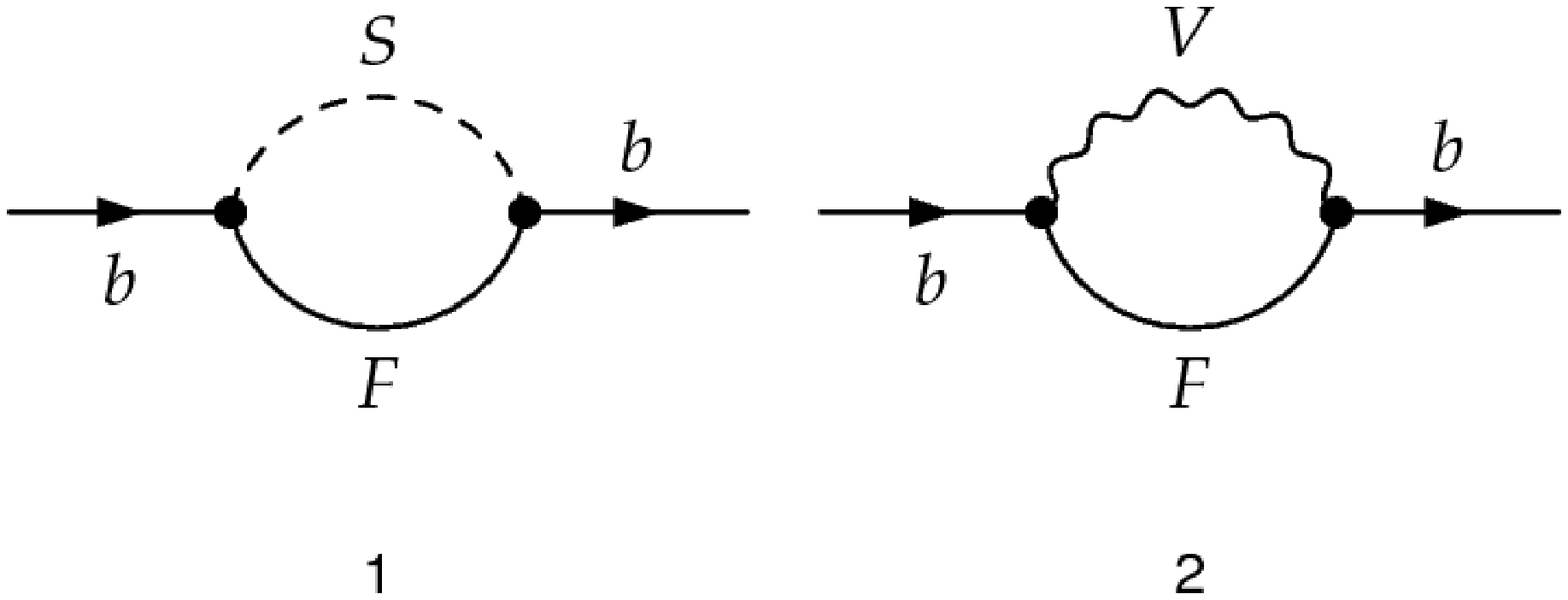, width=0.4\textwidth, angle=0}\\[10pt]
\epsfig{file=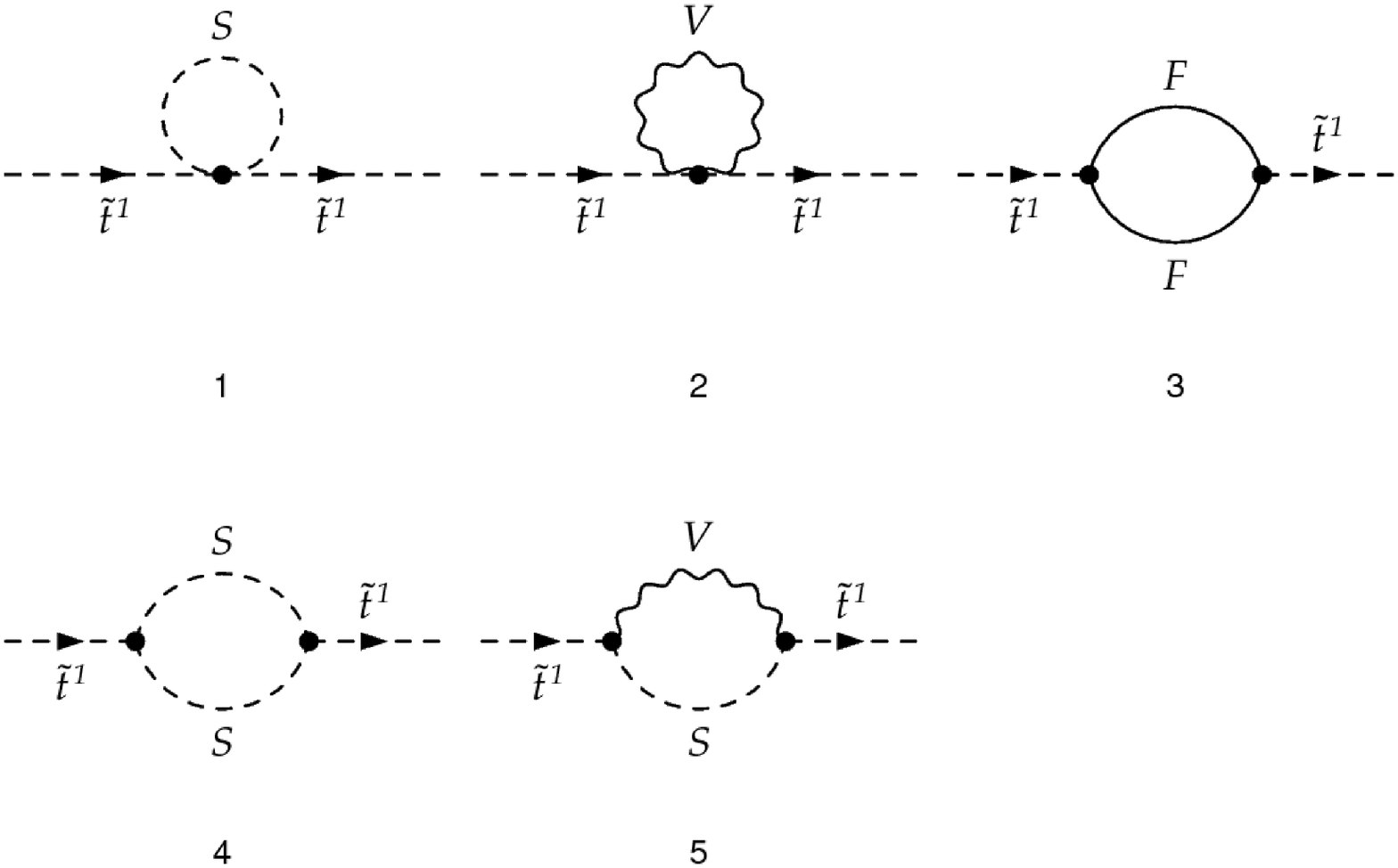, width=0.6\textwidth, angle=0}
\caption{Self-energy diagrams for chargino, bottom, and stop lines}
\label{fig:bgchistop_self}
\end{figure}

The contributions to the s-channel of the counter terms terms are, symbolically:
\bqa
N^{c.t.~s}_{1L}=
&&-~{g_s({\lambda^l\over2})\over s-m^2_b}\{
{3\over2}\delta Z^b_{L}A^{L}_i(\tilde{t}_a)
+{1\over2}\sum_{a'}\delta Z^*_{a'a}A^{L}_i(\tilde{t}_{a'})\nonumber\\
&&+\delta A^L_i(\tilde{t}_a)
+{1\over2}
\sum_{j}\delta \chi^{L}_{ji}A^{L}_j(\tilde{t}_a)\}  \\
%
%
N^{c.t.~s}_{1R}=
&&-~{g_s({\lambda^l\over2})\over s-m^2_b}\{
{3\over2}\delta Z^b_{R}A^{R}_i(\tilde{t}_a)
+{1\over2}\sum_{a'}\delta Z^*_{a'a}A^{R}_i(\tilde{t}_{a'})\nonumber\\
&&+\delta A^R_i(\tilde{t}_a)
+{1\over2}
\sum_{j}\delta \chi^{R}_{ji}A^{R}_j(\tilde{t}_a)\}  \\
%
%
N^{c.t.~s}_{3L}=
&&-~{m_bg_s({\lambda^l\over2})\over s-m^2_b}\{
(\delta Z^b_{L}+{1\over2}\delta Z^b_{R})A^{R}_i(\tilde{t}_a)
+{1\over2}\sum_{a'}\delta Z^*_{a'a}A^{R}_i(\tilde{t}_{a'})
\nonumber\\
&&+\delta A^R_i(\tilde{t}_a)
+{1\over2}
\sum_{j}\delta \chi^{R}_{ji}A^{R}_j(\tilde{t}_a)~\}
-m_bN^{c.t.~s}_{1R}       \\
%
%
N^{c.t.~s}_{3R}=
&&-~{m_bg_s({\lambda^l\over2})\over s-m^2_b}\{
(\delta Z^b_{R}+{1\over2}\delta Z^b_{L})A^{L}_i(\tilde{t}_a)
+{1\over2}\sum_{a'}\delta Z^*_{a'a}A^{L}_i(\tilde{t}_{a'})
\nonumber\\
&&+\delta A^L_i(\tilde{t}_a)
+{1\over2}
\sum_{j}\delta \chi^{L}_{ji}A^{L}_j(\tilde{t}_a)~\}
-m_bN^{c.t.~s}_{1L}
\eqa
and from $b$ s.e. one gets ($\eta=+1,-1$ means $R,L$):
\bqa
N^{s.e.~s}_{1\eta}=
&&~g_s({\lambda^l\over2}){1\over(s-m^2_b)^2}
[A^{\eta}_i(\tilde{t}_a)
(s(\Sigma^b_{\eta}(s)+\delta Z^b_{\eta})
+m^2_b(\Sigma^b_{-\eta}(s)+\delta Z^b_{-\eta})\nonumber\\
&&
+2m^2_b(\Sigma^b_{S}(s)-~{1\over2}
(\delta Z^b_{\eta}+\delta Z^b_{-\eta})-~{\delta m_b\over m_b}]
\eqa
\bqa
N^{s.e.~s}_{3\eta}=
&&~g_s({\lambda^l\over2}){1\over(s-m^2_b)^2}
[A^{-\eta}_i(\tilde{t}_a)
(sm_b(\Sigma^b_{\eta}(s)+\delta Z^b_{\eta})
+sm_b(\Sigma^b_{-\eta}(s)+\delta Z^b_{-\eta}))\nonumber\\
&&
+m_b(s+m^2_b)(\Sigma^b_{S}(s)-~{1\over2}
(\delta Z^b_{\eta}+\delta Z^b_{-\eta})-~{\delta m_b\over m_b}]
-m_bN^{s.e.~s}_{1~-\eta}\eqa
For the u-channel c.t. we obtain:
\bqa
N^{c.t.~u}_{2L}=
&&2g_s({\lambda^l\over2})\{
{1\over2}\delta Z^b_{L}A^{L}_i(\tilde{t}_a)
({1\over u-m^2_{\tilde{t}_a}})
+\sum_{a'}\bar{\delta} Z_{a'a}A^{L}_i(\tilde{t}_{a'})
({1\over u-m^2_{\tilde{t}_{a'}}})\nonumber\\
&&
+{1\over2}\sum_{a'}\delta Z^*_{a'a}A^{L}_i(\tilde{t}_{a'})
({1\over u-m^2_{\tilde{t}_{a}}})
+\delta A^L_i(\tilde{t}_a)({1\over u-m^2_{\tilde{t}_{a}}})
\nonumber\\
&&
+{1\over2}
\sum_{j}\delta \chi^{L}_{ji}A^{L}_j(\tilde{t}_a)
({1\over u-m^2_{\tilde{t}_{a}}})~\}
\eqa
\bqa
N^{c.t.~u}_{2R}=
&&2g_s({\lambda^l\over2})\{
{1\over2}\delta Z^b_{R}A^{R}_i(\tilde{t}_a)
({1\over u-m^2_{\tilde{t}_a}})
+\sum_{a'}\bar{\delta} Z_{a'a}A^{R}_i(\tilde{t}_{a'})
({1\over u-m^2_{\tilde{t}_{a'}}})\nonumber\\
&&
+{1\over2}\sum_{a'}\delta Z^*_{a'a}A^{R}_i(\tilde{t}_{a'})
({1\over u-m^2_{\tilde{t}_{a}}})
+\delta A^R_i(\tilde{t}_a)({1\over u-m^2_{\tilde{t}_{a}}})
\nonumber\\
&&
+{1\over2}
\sum_{j}\delta \chi^{L}_{ji}A^{L}_j(\tilde{t}_a)
({1\over u-m^2_{\tilde{t}_{a}}})~\}
\eqa
and from $\tilde{t}_a$ s.e.:
\bqa
N^{s.e.~u}_{2\eta}=
&&-2g_s({\lambda^l\over2}){1\over u-m^2_a}
\sum_{a'}\bar v_c(\chi^-_i)[A^L_i(\tilde{t}_{a'})P_L+
A^R_i(\tilde{t}_{a'})P_R]u(b){\hat{\Sigma}_{a'a}(u)\over
u-m^2_{a'}}
\eqa
The renormalized self-energy $\hat{\Sigma}_{a'a}(u)$
is defined below. Following \cite{Djouadi:1996wt,Kraml:1996kz,Eberl:1996wa,Arhrib:2003rp} we have:
\bq
\delta Z_{ba}={2\Sigma_{ba}(m^2_a)
\over m^2_b-m^2_a} \qquad
\delta Z_{aa}=
-[{d\Sigma_{aa}(p^2)\over dp^2}]_{p^2=m^2_a}
\eq
These results allow to write the renormalized
stop self-energies as:
\bq
\hat{\Sigma}_{aa}(p^2)=\Sigma_{aa}(p^2)
-\Sigma_{aa}(m^2_a)
-(p^2-m^2_a)
[{d\Sigma_{aa}(p^2)\over dp^2}]_{p^2=m^2_a}
\eq
and for $a\neq b$
\bq
\hat{\Sigma}_{ba}(p^2)=\Sigma_{ba}(p^2)
+{p^2-m^2_b\over m^2_{\tilde t_b}-m^2_a}
\Sigma_{ba}(m^2_{\tilde t_a})
+{p^2-m^2_a\over m^2_a-m^2_b}
\Sigma^*_{ab}(m^2_b)
\eq
The renormalization condition on the mixing angle is defined \cite{Eberl:1996wa}
in order to ensure the finiteness of the squark vertices.
\bq
\delta \theta_t={\Sigma_{12}(m^2_1)+\Sigma_{21}(m^2_2)\over
2(m^2_1-m^2_2)}={1\over4}[\delta Z_{12}-\delta Z_{21}]
\eq
which gives the needed:
\bq
\delta R_{1L}=\delta R_{2R}=\delta\cos\theta_t
=-\sin\theta_t \delta \theta_t
~~~~
\delta R_{1R}=-\delta R_{2L}=\delta\sin\theta_t
=\cos\theta_t \delta \theta_t
\eq
The various cunter terms for the quarks and gauge bosons have the following
explicit  form in terms of self-energies; for b, t quark and gauge part:
\bq
\delta Z^b_L=\delta Z^t_L\equiv \delta Z_L=
-\Sigma^b_L(m^2_b)-m^2_b[\Sigma^{'b}_L(m^2_b)+\Sigma^{'b}_R(m^2_b)
+2\Sigma^{'b}_S(m^2_b)]
\eq
\bq
\delta Z^b_R=
-\Sigma^b_R(m^2_b)-m^2_b[\Sigma^{'b}_L(m^2_b)+\Sigma^{'b}_R(m^2_b)
+2\Sigma^{'b}_S(m^2_b)]
\eq
\bq
\delta m_b={m_b\over2}Re[\Sigma^b_L(m^2_b)+\Sigma^b_R(m^2_b)
+2\Sigma^b_S(m^2_b)]
\eq
\bq
\delta Z^W_1-\delta Z^W_2=~{\Sigma^{\gamma Z}(0)\over s_Wc_W M^2_Z}
\eq
\bq
\delta Z^W_2 = - \Sigma^{'\gamma\gamma}(0) 
+2{c_W\over s_W M^2_Z}\Sigma^{\gamma Z}(0) 
+{c^2_W\over s^2_W}[{\delta M^2_Z\over M^2_Z} - 
{\delta M^2_W\over M^2_W}]
\eq
\bq
\delta M^2_W=Re\Sigma^{WW}(M^2_W)~~~~~\delta M^2_Z=Re\Sigma^{ZZ}(M^2_Z)
\eq
while the couterterms for the gauge coupling lead to:
\bq
{\delta g\over g}=\delta Z^W_1-~{3\over2}\delta Z^W_2
\eq
\bq
{\delta A^L_i(\tilde{t}_a)\over A^L_i(\tilde{t}_a)}
={1\over A^L_i(\tilde{t}_a)}\{\delta R_{aL}A^L_i(\tilde{t}_L)
+R_{aL}\delta A^L_i(\tilde{t}_L)
+\delta R_{aR}A^L_i(\tilde{t}_R)
+R_{aR}\delta A^L_i(\tilde{t}_R)\}
\eq
\bq
{\delta A^R_i(\tilde{t}_a)\over A^R_i(\tilde{t}_a)}
={\delta R_{aL}\over R_{aL}}
+{\delta A^R_i(\tilde{t}_L)\over A^R_i(\tilde{t}_L)}
\eq
\bq
{\delta A^L_i(\tilde{t}_L)\over A^L_i(\tilde{t}_L)}
={\delta g\over g}+{\delta Z^{+}_{1i}\over Z^{+}_{1i}}
\eq 
\bq
{\delta A^L_i(\tilde{t}_R)\over A^L_i(\tilde{t}_R)}
={\delta g\over g}+{\delta Z^{+}_{2i}\over Z^{+}_{2i}}
+{\delta m_t\over m_t}-{\delta M_W\over M_W}
-{\delta \sin\beta\over \sin\beta}
\eq 
\bq
{\delta A^R_i(\tilde{t}_L)\over A^R_i(\tilde{t}_L)}
={\delta g\over g}+{\delta Z^{-*}_{2i}\over Z^{-*}_{2i}}
+{\delta m_b\over m_b}-{\delta M_W\over M_W}
-{\delta \cos\beta\over \cos\beta}
\eq 
From \cite{Wan:2001rt} we have for $\delta\tan\beta\over\tan\beta$:
\bq
{\delta\tan\beta\over\tan\beta}={Re\Sigma_{H^+W^+}(m^2_{H^+})
\over M_W\sin2\beta}
\eq
\bq
{\delta\sin\beta\over\sin\beta}=\cos^2\beta
{\delta\tan\beta\over\tan\beta}~~~~~~
{\delta\cos\beta\over\cos\beta}=-\sin^2\beta
{\delta\tan\beta\over\tan\beta}
\eq
We need also the counterterms for the chargino mixing
matrices. Applying the method of  \cite{Eberl:1996wa} \cite{Wan:2001rt} 
requiring the cancellation of the antihermitean part
of the wave function renormalization, we have:
\bq
\delta Z^+_{1i}={1\over4}\sum_k Z^+_{1k} 
(\delta\chi^{L*}_{ik}-\delta\chi^{L}_{ki})
\eq
\bq
\delta Z^+_{2i}={1\over4}\sum_k Z^+_{2k} 
(\delta\chi^{L*}_{ik}-\delta\chi^{L}_{ki})
\eq
\bq
\delta Z^{-*}_{2i}={1\over4}\sum_k Z^{-*}_{2k} 
(\delta\chi^{R*}_{ik}-\delta\chi^{R}_{ki})
\eq
Using the $\chi^+$ chargino c.t. and s.e.
$\delta\chi^{L,R}_{ij}$ listed below
the counter terms for $\chi^+_i$ are obtained from
the field transformation:
\bq
\chi^+_i\to (1+{1\over2}
[\delta\chi^L_{ij}P_L+\delta\chi^R_{ij}P_R])\chi^+_j
\eq
They are obtained  by applying the method proposed in \cite{Wan:2001rt} \cite{Denner:1990yz}
\cite{Kniehl:1996bd}  and in terms of the $j\to i$ bubble of momentum $p$ they read:
\bq
\Sigma_{ij}=\psl P_L \Sigma^L_{ij}
+\psl P_R \Sigma^R_{ij}+P_L \Sigma^S_{ij}
+P_R \Sigma^{\bar S}_{ij}
\eq
with $\Sigma^{\bar S}_{ij}=\Sigma^{S*}_{ji}$ and:
%
%
%
\bq
\delta\chi^L_{ii}=-\{\Sigma^{L}_{ii}(M^2_i)
+M^2_i[\Sigma^{'L}_{ii}(M^2_i)+\Sigma^{'R}_{ii}(M^2_i)]
+M_i[\Sigma^{'S}_{ii}(M^2_i)+\Sigma^{'\bar S}_{ii}(M^2_i)]\}
\eq
\bq
\delta\chi^R_{ii}=-\{\Sigma^{R}_{ii}(M^2_i)
+M^2_i[\Sigma^{'L}_{ii}(M^2_i)+\Sigma^{'R}_{ii}(M^2_i)]
+M_i[\Sigma^{'S}_{ii}(M^2_i)+\Sigma^{'\bar S}_{ii}(M^2_i)]\}
\eq
and for $i\neq j$ 
\bq
\delta\chi^L_{ij}={2\over M^2_i-M^2_j}
\{M^2_j\Sigma^{L}_{ij}(M^2_j)
+M_iM_j\Sigma^{R}_{ij}(M^2_j)
+M_i\Sigma^{S}_{ij}(M^2_j)+M_j\Sigma^{\bar S}_{ij}(M^2_j)\}
\eq
\bq
\delta\chi^R_{ij}={2\over M^2_i-M^2_j}
\{M^2_j\Sigma^{R}_{ij}(M^2_j)
+M_iM_j\Sigma^{L}_{ij}(M^2_j)
+M_j\Sigma^{S}_{ij}(M^2_j)+M_i\Sigma^{\bar S}_{ij}(M^2_j)\}
\eq
All the self-energy functions $\Sigma$ are computed from the diagrams of Fig.~\ref{fig:bgchistop_self}.

\newpage

\hfill

\newpage

\hfill

\hfill

\hfill

\hfill

\hfill

\hfill

\hfill

\hfill

\hfill


\end{document}